\documentclass[aps,prb,twocolumn,secnumarabic,superscriptaddress]{revtex4-1}

\usepackage{makecell}
\usepackage{array}
\newcommand{\ra}[1]{\renewcommand{\arraystretch}{#1}} 
\newcolumntype{"}{@{\hskip\tabcolsep\vrule width 1pt\hskip\tabcolsep}}

\usepackage{amsmath}
\usepackage{amssymb}
\usepackage{amsbsy}
\usepackage{wasysym}

\usepackage{titlesec}
\usepackage{bm}
\usepackage{comment}
\usepackage{verbatim}

\usepackage{graphicx}
\usepackage{subfigure}
\usepackage{tabularx}

\usepackage{color}
\usepackage[colorlinks,bookmarks=false,citecolor=blue,linkcolor=red,urlcolor=blue]{hyperref}

\definecolor{darkred}{rgb}{0.7,0.0,0.0}

\definecolor{darkblue}{rgb}{0,0.02,0.45}
\def\cdbl{\color{darkblue}}
\definecolor{darkgreen}{rgb}{0.02,0.45,0.0}

\definecolor{violet}{rgb}{0.8,0.2,0.6}

\def\be{\begin{equation}}
\def\ee{\end{equation}}
\def\bea{\begin{eqnarray}}
\def\eea{\end{eqnarray}}

\def\mc{\mathcal}

\usepackage{times}

\begin{document}
\date{\today}

\title{Microscopic theory of the nearest-neighbor valence bond sector \\ of the spin-1/2 kagome antiferromagnet}

\author{Arnaud Ralko}\email{arnaud.ralko@neel.cnrs.fr}
\affiliation{Institut N\'eel, UPR2940, Universit\'e Grenoble Alpes et CNRS, Grenoble, FR-38042 France}

\author{Fr\'ed\'eric Mila}\email{frederic.mila@epfl.ch}
\affiliation{Institute of Theoretical Physics, \'Ecole Polytechnique F\'ed\'erale de Lausanne, CH-1015 Lausanne, Switzerland}

\author{Ioannis Rousochatzakis}\email{irousoch@umn.edu}
\affiliation{School of Physics and Astronomy, University of Minnesota, Minneapolis, MN 55455, USA}

\date{\today}

\begin{abstract}
The spin-1/2 Heisenberg model on the kagome lattice, which is closely realized in layered Mott insulators such as ZnCu$_3$(OH)$_6$Cl$_2$, is one of the oldest and most enigmatic spin-1/2 lattice model. While the numerical evidence has accumulated in favor of a quantum spin liquid, the debate is still open as to whether it is a $Z_2$ spin liquid with very short-range correlations (some kind of Resonating Valence Bond spin liquid), or an algebraic spin-liquid with power-law correlations. To address this issue, we have pushed the program started by Rokhsar and Kivelson in their derivation of the effective quantum dimer model description of Heisenberg models to unprecedented accuracy for the spin-1/2 kagome, by including all the most important virtual singlet contributions on top of the orthogonalization of the nearest-neighbor valence bond singlet basis. Quite remarkably, the resulting picture is a competition between a $Z_2$ spin liquid and a diamond valence bond crystal with a 12-site unit cell, as in the DMRG simulations of Yan, Huse and White. Furthermore, we found that, on cylinders of finite diameter $d$, there is a transition between the $Z_2$ spin liquid at small $d$ and the diamond valence bond crystal at large $d$, the prediction of the present microscopic description for the 2D lattice. These results show that, if the ground state of the spin-1/2 kagome antiferromagnet can be described by nearest-neighbor singlet dimers, it is a diamond valence bond crystal, and, a contrario, that, if the system is a quantum spin liquid, it has to involve long-range singlets, consistent with the algebraic spin liquid scenario.
\end{abstract}

\maketitle

\section{Introduction}
\begin{figure}[!t]
\vspace{-0.25cm}
\includegraphics[width=0.45\textwidth,clip]{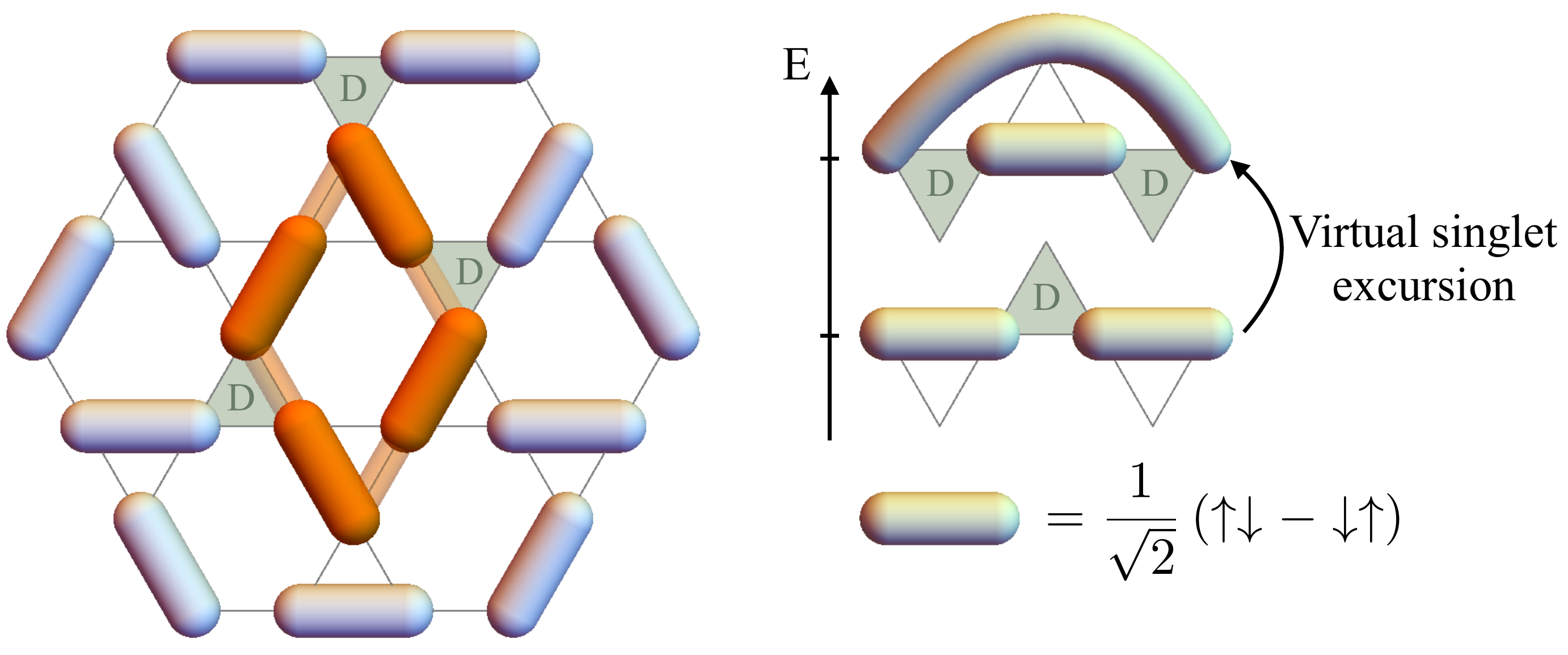}
\vspace{-0.25cm}
\caption{A dimer covering of the kagome lattice, where each dimer represents a spin singlet between NN spins. Tunneling processes shift the dimers around closed loops (orange). Such processes are mediated by longer-range singlets that are virtually excited around empty, or `defect' triangles (D). The ensuing amplitude depends on the particular dimer environment outside the loop.~\cite{IoannisZ2,L4L8}}\label{fig:Model}
\end{figure}

\vspace*{-0.3cm}
The idea that spins in a solid can evade ordering down to zero temperature by forming a correlated quantum spin liquid (QSL) has a very long history.~\cite{Anderson1973,FazekasAnderson74,Anderson1987,LiangDoucotAnderson88,Sachdev92,Sandvik05,HFMBook,Balents2010} 
Such phases host topological properties, long-range entanglement, and fractionalized excitations with anyonic statistics, and have been discussed for applications in quantum computing.~\cite{Nayak2008,Dennis2002,Freedman2002,Kitaev2003}
One of the simplest models that has long been predicted~\cite{Sachdev92} to host a QSL phase is the antiferromagnetic (AF) nearest-neighbor (NN) spin-1/2 Heisenberg model on the kagome lattice (Fig.~\ref{fig:Model}), described by the spin Hamiltonian \be\label{eq:Hheis} \mc{H}=J\sum\nolimits_{\langle ij\rangle} {\bf S}_i\cdot{\bf S}_j\,, \ee where ${\bf S}_i$ and ${\bf S}_j$ are NN spins on the lattice, and $J\!>\!0$ is the exchange coupling. This model, which describes closely the layered ZnCu$_3$(OH)$_6$Cl$_2$,~\cite{Shores05,Han2012,Norman2016} has been studied intensively in the last 30 years with a multitude of techniques,~\cite{
Elser1989,ZengElser1990,Chalker1992,LeungElser1993,Lecheminant1997,Waldtmann1998,Sindzingre2000,Laechli2011,Nakano2011,
ZengElser95,Mila1998,MambriniMila2000,Misguich02,Misguich03,Schwandt2010,Poilblanc2010,IoannisZ2,Hao14,
Ran2007,Iqbal2011a,Iqbal2011b,Tay2011,Iqbal2013,Clark2013,Iqbal2014,
Gotze2011,
Budnik2004,Capponi2004,Capponi2013,
SinghHuse1992,SinghHuse07,
Jiang2008,YanHuseWhite2011,Shollwock2012,He2016,
Vidal2010,Xie2014,Liao2017,Mei2017} 
%
and with conclusions that go in all possible directions. At that stage, the problem is no longer to find the right solution, but to eliminate the wrong ones! 

Now, even concentrating on the most recent (hence arguably most reliable) numerical results, the situation is still debated. A breakthrough from density matrix renormalization group (DMRG) studies~\cite{YanHuseWhite2011,Shollwock2012,Jiang2012} showed evidence for a gapped $Z_2$ QSL, proposed by Anderson in 1973.~\cite{Anderson1973} 
This conclusion is challenged in a more recent DMRG study,~\cite{He2016} with evidence for a gapless, U(1) spin liquid, which would be in line with variational Monte Carlo studies.~\cite{Iqbal2013,Iqbal2014}
A recent application~\cite{Liao2017} of tensor-network methods delivers also a gapless spin liquid, but this is challenged by another tensor network study~\cite{Mei2017}  supporting the $Z_2$ liquid scenario.
These conflicting results all the more demonstrate that this is a paradigmatic strongly correlated problem with many orders that compete at tiny energy scales, as highlighted explicitly by a full diagonalization tour de force.~\cite{Laeuchli2016}

In view of this difficulty, we adopt the point of view that, for powerful that they might be, numerical simulations need to be complemented by microscopic and/or analytical approaches to identify the degrees of freedom at play, their interaction, and the resulting physics. This means in particular that, if the system is a $Z_2$ spin liquid or some kind of valence bond crystal (VBC), it has to be possible to derive an effective model for the singlet sector in terms of nearest-neighbour valence bond (NNVB) dimers whose properties can be compared to the numerical simulations that have identified this kind of physics. This goes back to the pioneering work of Rokhsar and Kivelson,~\cite{RokhsarKivelson} who suggested to describe the resonating valence bond (RVB) physics in terms of a quantum dimer model (QDM), by restricting the original SU(2) spin Hamiltonian to the NNVB basis. But since the NNVB configurations are not orthogonal for SU(2) spins, the first step towards a QDM description is to orthogonalize the basis. This task, which was initially performed to second order in the overlap integral by Rokhsar and Kivelson,~\cite{RokhsarKivelson} has been tackled in many subsequent works,~ \cite{ZengElser95,MambriniMila2000,Misguich02,Misguich03} and more recently has been performed to very high order,~\cite{Schwandt2010,Poilblanc2010} see also related discussion in Ref.~[\onlinecite{IoannisZ2}].

However, even if the orthogonalization is done essentially exactly, the restriction of the Hamiltonian to the NNVB basis amounts to only the first-order contribution in degenerate perturbation theory, and, to go beyond, one has to include higher-order virtual excitations outside the basis. This program has only started recently~\cite{IoannisZ2,L4L8}, with the conclusion that these virtual singlet fluctuations change the amplitude of some QDM processes strongly enough to change the physics.

In the present paper, we go one step further in the case of the kagome antiferromagnet by considering the most accurate QDM considered so far, in which the dependence of various processes on the embedding is taken into account for the first time when studying the properties of the QDM. The resulting picture, a competition between a $Z_2$ spin liquid and a diamond-like VBC is quite different from the properties of more elementary QDM descriptions, but it agrees with the DMRG results of Yan {\it et al}.~\cite{YanHuseWhite2011} We take this as a strong evidence that our QDM description is accurate, and that the present investigation to a large extent completes the QDM approach to the spin-1/2 kagome antiferromagnet that started more than twenty years ago.~\cite{ZengElser95}

Of course, this is not the final word about the kagome antiferromagnet. By construction, our QDM only includes fluctuations involving finite-range singlets, and if the low-energy physics is controlled by long-range singlets, it cannot be described by our QDM. In that respect, it would be very nice to have a physical picture of the competing algebraic phases in a similar language, but with long-range singlets. 

The remaining part of the article is organized as follows. Sec.~\ref{sec:MainResults} summarizes our most crucial results. Sec.~\ref{sec:RVB} describes the microscopic derivation of the QDM parameters and their embedding dependence, along with the identification of the most important ingredients. Secs.~\ref{sec:QDMresults} and \ref{sec:NotIncludedTerms} provide our detailed numerical calculations of the QDM, the physical origin of the diamond VBC and its accompanying low-lying excitations, as well as the individual role of the various tunneling terms. In Sec.~\ref{sec:GSEnergy} we benchmark our results by comparing the ground state energy delivered from the effective description to published ED data on the Heisenberg model. Sec.~\ref{sec:DMRG} gives our numerical QDM calculations on asymmetric tori that resemble the cylinder geometries of DMRG, and show that the behaviors at small and large cylinder diameters $d$ are qualitatively different. The discussion of Sec.~\ref{sec:Disc} gives a broader perspective of our results. We finally provide five Appendices with various technical details and auxiliary information.

\vspace*{-0.3cm}
\section{Summary of main results}\label{sec:MainResults}
\vspace*{-0.3cm}
The first crucial result emerging from the microscopic RVB description presented below is that only 5\% of the energy arises from tunneling. The remaining 95\% arises from potential energy contributions that are essentially the same for all NNVB states. This explains  why there are so many different orders that compete at tiny energy scales.

Second, the short-range tunneling physics is essentially governed only by the dimer resonances around loops of length $L\!=\!8$, and, in particular, by the variations of the corresponding tunneling amplitudes with the valence bond pattern outside the loop. This `lattice embedding' effect is one of the main ramifications of virtual singlets and derives from their long-range nature.~\cite{IoannisZ2,L4L8}

Third, changing artificially the degree of this embedding dependence drives the system through a quantum critical point, which separates the two competing phases discussed in Ref.~[\onlinecite{YanHuseWhite2011}], the gapped $Z_2$ QSL and the so-called diamond valence bond crystal (VBC). 
In contrast to previous proposals,~\cite{Wan2013,Hao14,Hwang2015} this crystal is stabilized here by tunneling and not by potential energy. The physical tunneling parameters, as extracted from a cluster exact diagonalization (ED) method, place the system in the VBC side of the critical point and not in the spin liquid side. 

Fourth, the apparent discrepancy with previous DMRG calculations that are in favor of a gapped $Z_2$ QSL~\cite{YanHuseWhite2011,Shollwock2012,Jiang2012} can be resolved by checking explicitly the influence of boundary effects in finite-size calculations using clusters with different geometries. The results for the topological gap reveal that, on cylinders with fixed diameter $d$, there is a phase transition between the $Z_2$ spin liquid at small $d$ and the diamond VBC at large $d$, the prediction of the present microscopic RVB description for the 2D lattice.

Fifth, the microscopic description gives important insights for the excitation spectrum as well. 
One can identify domain-wall excitations that separate different orientations of the diamond VBC, as well as vortices associated with the intersection of three domain walls. The energy cost of these excitations is controlled directly by the variations in the tunneling amplitudes with the lattice embedding, showing that the melting of the diamond crystal toward the $Z_2$ spin liquid proceeds via a condensation of domain walls and vortices.
The proximity of the diamond VBC to the critical point then implies a sub-extensive number of low-lying domain-walls and an extensive number of low-lying vortices, which can partly account for the high density of low-lying singlets found previously by ED.~\cite{Lecheminant1997,Waldtmann1998,Nakano2011,Laechli2011,Laeuchli2016}

It is finally established that resonances with $L\!\neq\!8$ do not affect the physics in any appreciable way, either because they are too weak or due to significant phase space constraints. 
In addition, the $Z_2$ QSL is adiabatically connected to the Rokhsar-Kivelson wavefunction of the integrable models of Misguich {\it et al}~\cite{Misguich02} and Hao {\it et al},~\cite{Hao14} which is non-trivial because the parameters of these models are far from the microscopic ones. 
These aspects signify a qualitative reduction in the complexity of the problem, at the heart of which lies the role of virtual singlets, in conjunction with the phase space constraints mentioned above. 

In the following, we set out to describe this complexity reduction and present our numerical results from ED on the Heisenberg model, ED on the microscopic quantum dimer model (QDM), as well as Green's function Monte Carlo (GFMC) on a minimal QDM with $L\!=\!8$ resonances only, which has no negative sign problem. 

\begin{figure}[!t]
\vspace{-0.25cm}
\includegraphics[width=0.45\textwidth,clip]{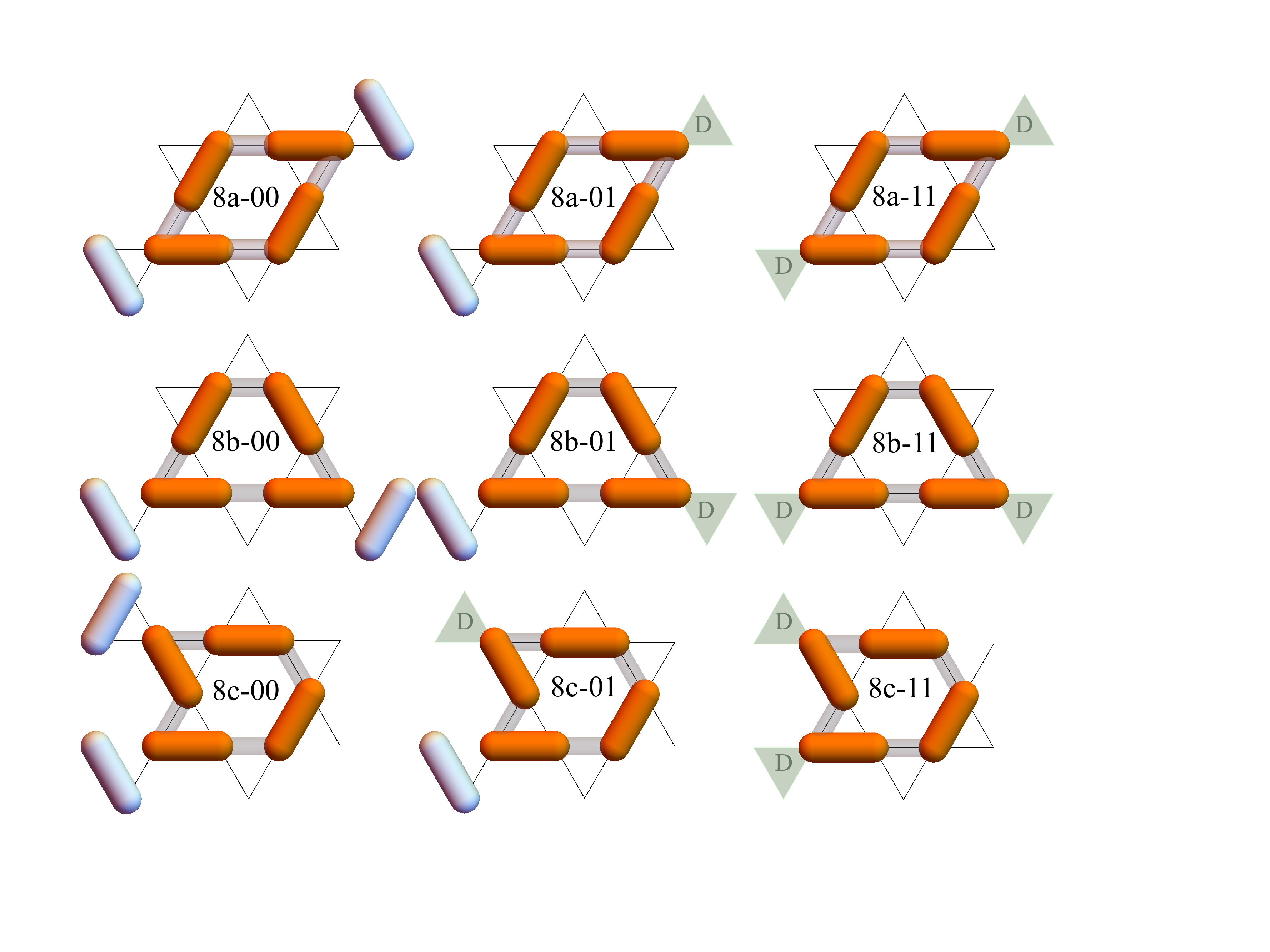}
\vspace{-0.25cm}
\caption{The most relevant, loop-8 tunneling processes of the microscopic RVB model. Different rows give the three topologically different loops, and different columns designate the dependence of the amplitudes on the possible presence of extra defect triangles nearby.}\label{fig:L8}
\end{figure}

\vspace*{-0.3cm}
\section{Effective RVB description}\label{sec:RVB}
\vspace*{-0.3cm}
Ideally, magnetically disordered phases with finite correlation lengths are described in terms of short-range valence bond coverings, where nearby spins organize into singlet pairs or valence bonds.~\cite{Anderson1973,FazekasAnderson74,Anderson1987,LiangDoucotAnderson88,Sandvik05} The kagome is special in that singlets of the shortest possible range, i.e. between NN spins, form states with a huge variational energy lead of at least $3J/2$ compared to states with longer-range singlets.~\cite{Elser1989,ZengElser95} So NNVB states form an excellent variational starting basis.

The dynamics in this basis is then cast in terms of tunneling processes (plus potential terms) between different NNVB states,~\cite{ZengElser95,MambriniMila2000,Misguich02,Misguich03,Schwandt2010,Poilblanc2010,IoannisZ2} see Fig.~\ref{fig:Model}. The most local (and thus important) events connect NNVB states by shifting the singlets around a loop that encircles a single hexagon. There are eight topologically distinct loops of this type, with lengths $L\!=\!6$ (one process), $8$ (three processes, `8a', `8b' and `8c'), $10$ (three processes, `10a', `10b' and `10c') and $12$ (one process).~\footnote{We note that `8a', `8b', `8c', `10a', `10b' and `10c' correspond, respectively, to the processes `C1', `C3', `C2', `B1', `B3', and `B2' of Ref.~[\onlinecite{IoannisZ2}].} For each type there is a tunneling amplitude ($t_6$, $t_{8a}$, etc) and a potential energy term ($V_6$, $V_{8a}$, etc).

Calculating these parameters from (\ref{eq:Hheis}) has been a technical challenge for many years, partly due to the non-orthogonality of the NNVB states.~\cite{RokhsarKivelson,ZengElser95} While this problem is now largely resolved,~\cite{Schwandt2010,Poilblanc2010,IoannisZ2} a more fundamental problem is the correct treatment of the virtual longer-range singlet fluctuations (Fig.~\ref{fig:Model}), as emphasized in the early work of Zeng and Elser.~\cite{ZengElser95} Mathematically, the problem amounts to going beyond the first-order variational projection of (\ref{eq:Hheis}) into the NNVB basis, and include higher-order terms. 

The simple idea of Ref.~[\onlinecite{IoannisZ2}] is that since the effective parameters are governed by local processes, one can extract them from the exact spectra of specially designed Heisenberg clusters, in analogy to how one extracts e.g.\! the Heisenberg exchange $J$ from the exact solution of a two-site Hubbard model. What's more, by systematically enlarging the size of the clusters one can incorporate the renormalization effect from virtual singlets of longer and longer range $R$, and establish the convergence with $R$ (see Ref.~[\onlinecite{IoannisZ2}] for a precise definition of $R$).

\vspace*{-0.3cm}
\subsection{The `R2' model}\vspace*{-0.3cm}
The effective parameters extracted in Ref.~[\onlinecite{IoannisZ2}] at the $R\!=\!2$ level are, in units of $J$: 
$t_6\!=\!0.146$, $t_{8a}\!=\!-0.082$, $t_{8b}\!=\!-0.084$, $t_{8c}\!=\!-0.067$, $t_{10a}\!=\!0.048$, $t_{10b}\!=\!0.025$, $t_{10c}\!=\!0.040$, $t_{12}\!=\!0$, 
while from the potential terms, only $V_6\!=\!0.097$ is appreciable. The loop-six terms have not yet converged at $R\!=\!2$, and the above numbers for $t_6$ and $V_6$ are upper bounds. Here, we take the values $t_6\!=\!0.127$ and $V_6\!=\!0.073$ obtained from a slightly larger cluster that is intermediate between $R\!=\!2$ and $R\!=\!3$ (see App.~\ref{app:HC6}).
Importantly, this uncertainty in $t_6$ does not eventually matter, as shown below.
Note also that the difference between $t_{8a}$, $t_{8b}$ and $t_{8c}$ (and similarly for loop-10 processes) is missed by the first-order truncation to the NNVB basis and is  one of the qualitative effects of virtual singlets.~\cite{IoannisZ2}

\begin{table*}[!t] \ra{1.1}
\begin{ruledtabular}
\caption{Tunneling amplitudes of the 9 loop-8 and 29 loop-10 processes of the `R2+E' model. The remaining two processes are the tunneling and potential terms of loop-six processes: $t_6\!=\!0.127$ and $V_6\!=\!0.097$, the same as in the `R2' model. All numbers are in units of $J$. 
The different loop-8 processes are shown in Fig.~\ref{fig:L8} and in Fig.~\ref{fig:loop8} of App.~\ref{app:HC810}. The loop-10 are shown in Figs.~\ref{fig:loop10a}-\ref{fig:loop10c} of App.~\ref{app:HC810}. There we also provide the list of Heisenberg clusters from which the amplitudes are extracted. 
The numbers in the first column coincide with the parameters of the `R2' model. 
Note that the numbers for `8a-00', `8a-01' and `8a-11' were already given in the Supplementing Material of Ref.~[\onlinecite{IoannisZ2}].
}\label{tab:tR2+E}
\begin{tabular}{cccccccccccc}
{\cdbl8a-00} & {\cdbl8a-01} & {\cdbl8a-11} & {\cdbl8b-00}  & {\cdbl8b-01}  & {\cdbl8b-11}  & {\cdbl8c-00} & {\cdbl8c-01} & {\cdbl8c-11} &&&\\
-0.082462 & -0.059786 & -0.041389 & -0.084103& -0.065086 & -0.055111 & -0.067355 & -0.04769 & -0.031227 &&&\\
\hline
{\cdbl10a-0000} & {\cdbl10a-1000}  & {\cdbl10a-1100} & {\cdbl10a-1010} & {\cdbl10a-1001} & {\cdbl10a-1110} & {\cdbl10a-1111} & & &&&\\
0.04779 & 0.039769 & 0.031067 & 0.03589 & 0.036154 & 0.030841  & 0.029065 &&&&&\\
\hline
{\cdbl10b-0000} & {\cdbl10b-1000}  & {\cdbl10b-0100} & {\cdbl10b-1100} & {\cdbl10b-1010} & {\cdbl10b-1001} &  {\cdbl10b-0110} & {\cdbl10b-1110} & {\cdbl10b-1101} & {\cdbl10b-1111} &  &\\ 
0.024966 & 0.01412 & 0.01455 & 0.00588 & 0.004937 & 0.00506 & 0.005938 & 0.001356 & 0.001633 & 0.005741 &&\\
\hline
{\cdbl10c-0000} & {\cdbl10c-1000}  & {\cdbl10c-0100} & {\cdbl10c-0010} & {\cdbl10c-1100} & {\cdbl10c-1010} &  {\cdbl10c-0110} & {\cdbl10c-0101} & {\cdbl10c-1110} & {\cdbl10c-1101} & {\cdbl10c-0111} &{\cdbl10c-1111}\\ 
0.03957 & 0.037377 & 0.027745 & 0.028579 & 0.028716 & 0.029442 & 0.017737 & 0.016515 & 0.020661 & 0.019595 & 0.008139 & 0.012273
\end{tabular}
\end{ruledtabular}
\end{table*}

\vspace*{-0.3cm}
\subsection{The `R2+E' model}\vspace*{-0.3cm}
Another qualitative effect of virtual singlets~\cite{IoannisZ2,L4L8} is the fact that the amplitudes depend not only on the type of loop but also on the particular NNVB environment `E' away from the loop (see Fig.~\ref{fig:Model}), i.e. $t_{8a}$ should be replaced with $t_{8a\text{-E}}$, etc. This adds another layer of complexity because it effectively increases the number of parameters. As it turns out however, it is in this extra layer of complexity that the crucial physical insights lie.

Let us take, for example, the `8a' process of Fig.~\ref{fig:L8} and restrict ourselves to the possible NNVB configurations on the two triangles that share a single site with the loops. Each of these triangles can either have a singlet or be empty. The latter possibility comes with enhanced quantum fluctuations (and therefore a different tunneling amplitude), because these so-called `defect triangles' do not satisfy the Hamiltonian locally.~\cite{Elser1989} Altogether, we get three possible nearby environments, see first row of Fig.~\ref{fig:L8}:  `8a-00' for the case where neither of the two triangles is empty, `8a-01' when one is empty, and `8a-11' when both are empty. Similarly, we get three nearby environments for `8b' and three for `8c', see Fig.~\ref{fig:L8}. 
Likewise, there is one nearby environment for loop-six, seven for `10a', ten for `10b', and twelve for `10c'.  Altogether, this increases the number of most relevant parameters from 8 to 40, and leads to the `R2+E' model. The values of the parameters are provided in Table~\ref{tab:tR2+E}. 

Naturally, the number of parameters increases further by considering more distant triangles. In turn, this induces further (but much weaker) indentations, but the essential physics is already revealed at the `R2+E' level, as shown below.

\vspace*{-0.3cm}
\subsection{The minimal, loop-8 model}\label{sec:minimal}
\vspace*{-0.3cm}
We next examine how far is the above microscopic RVB parameters from the integrable QSL models of Misguich {\it et al}~\cite{Misguich02} and Hao {\it et al}.~\cite{Hao14} These models share the same ground state, namely the equal amplitude superposition of all NNVB states, within a given topological sector.
In the former model, all tunneling amplitudes are equal to $-1$ and all potential terms vanish. These values are very different, both in magnitude and in relative signs, from `R2+E'. 
In the model by Hao {\it et al},~\cite{Hao14} all parameters vanish except $(t_8,V_8)\!=\!(-1,1)$, which also appear to be far from `R2+E'. In particular, the large $V_8$ cannot be accounted for by the microscopic model.
However, $V_8$ is not essential for spin liquidity because the liquid phase includes the point $(t_8,V_8)\!=\!(-1,0)$,~\cite{Hao14}  see also Ref.~[\onlinecite{AndreasQDM}]. 
In the following we set out to show that this special  `T8' point is in fact closer to the microscopic model than what is expected at first sight.

To this end, we consider a simplified version of the `R2+E' model, where we keep only the nine loop-8 processes of Fig.~\ref{fig:L8}, which will be referred to as the `T8+VS' model, where `VS' stands for the effect of virtual singlets. To probe this effect explicitly, we introduce a parameter $x$ to interpolate between the `T8' model ($x\!=\!0$) and the `T8+VS' model ($x\!=\!1$). Namely, we will consider the minimal QDM Hamiltonian 
\be\label{eq:Hsimpl}
\mc{H}_{\text{m}}(x) = (1\!-\!x)~\mc{H}_{\text{T8}} + x ~\mc{H}_{\text{T8+VS}}~,
\ee
and then later reinstate the remaining terms from `R2+E'. The physical parameters for the loop-eight processes are the ones corresponding to $x\!=\!1$.

\vspace*{-0.3cm}
\section{Results for the minimal model}
\label{sec:QDMresults}
\begin{figure*}[!t]
\vspace{-0.25cm}
\includegraphics[width=0.99\textwidth,clip]{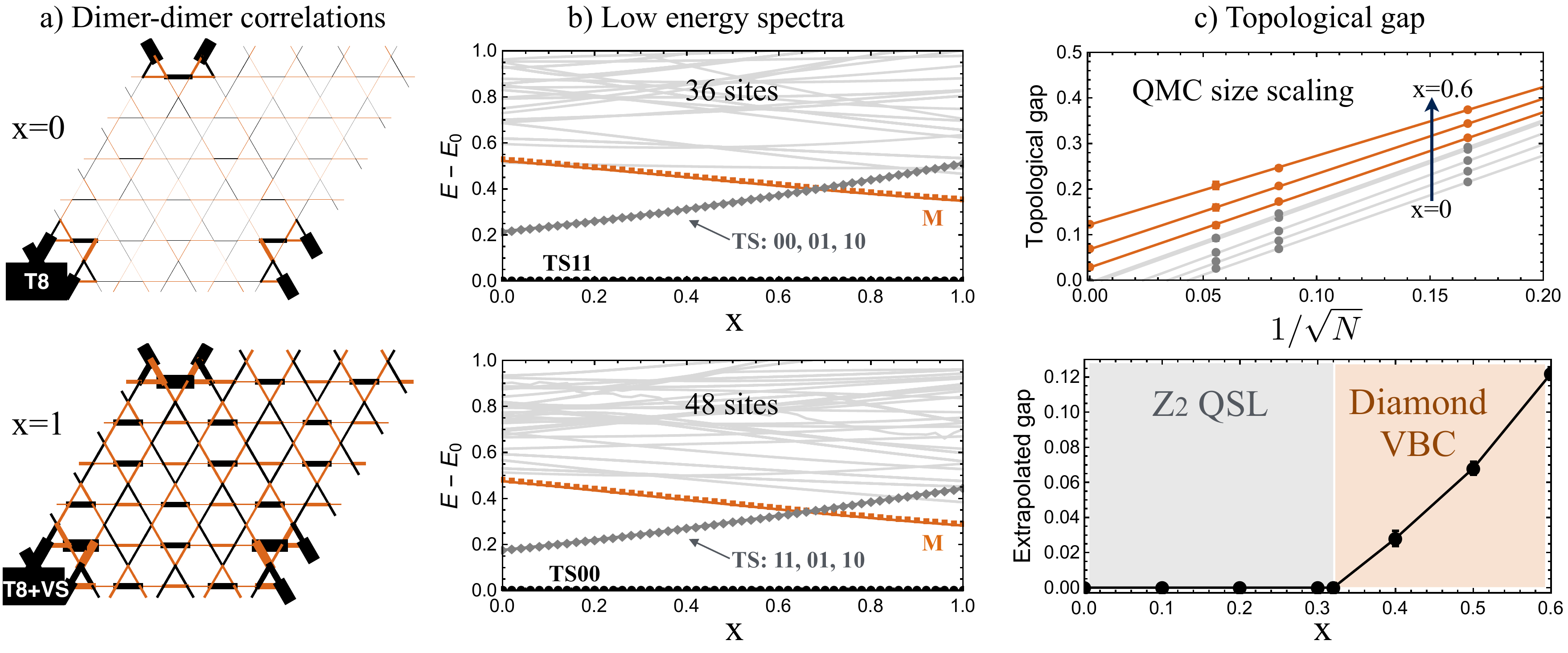}
\vspace{-0.25cm}
\caption{Numerical results for the minimal model $\mc{H}_{\text{m}}(x)$ of Eq.~(\ref{eq:Hsimpl}). 
(a) Connected dimer-dimer correlations in the ground state of the 108-site cluster for the `T8' ($x\!=\!0$) and the `T8+VS' ($x\!=\!1$) model. The thick bond on the lower left corner is the reference bond. Positive (negative) correlations are shown by black (red) color.
(b) Low-energy spectra as a function of $x$ measured from the ground state energy $E_0$ and in units of $t_6\!=\!0.127J$. The lowest energy states of the four topological sectors are denoted by `TS00', `TS01', `TS10' and `TS11'. The level indicated by `{\bf M}' is the lowest state with momentum at the middle of the Brillouin-zone edge. 
(c) Topological gap (in units of $t_6\!=\!0.127J$) as calculated from Green's function Monte Carlo up to 324 sites for various control parameters $x$. The gap scales well as $1/\sqrt{N}$. The lower panel shows the extrapolated value at the thermodynamic limit. 
The geometry of the clusters is discussed in App.~\ref{App:1}.
}\label{fig:SimplifiedModelResults}
\end{figure*}

\vspace*{-0.3cm}
The numerical results for the dimer-dimer correlations (ED on 108 sites), the low-energy spectra (ED on 36- and 48-site clusters), and the topological gap (GFMC up to 324 sites), are shown in Fig.~\ref{fig:SimplifiedModelResults}. The results demonstrate that $\mc{H}_{\text{m}}(x)$ hosts two main competing states, the $Z_2$ spin liquid at small $x$, and the diamond VBC at large $x$, which are the two states reported by Yan {\it et al}.~\cite{YanHuseWhite2011}

The connected dimer-dimer correlations are defined as 
\be
\langle \psi| {\bf D}_{ij} {\bf D}_{kl}  |\psi\rangle
-\langle \psi| {\bf D}_{ij} |\psi\rangle \langle \psi|{\bf D}_{kl}|\psi\rangle\,,
\ee 
where $|\psi\rangle$ is the ground state of $\mc{H}_{\text{m}}(x)$  in the given cluster, and ${\bf D}_{ij}$ and ${\bf D}_{kl}$ are dimer operators on the nearest-neighbor bonds of sites $(i,j)$ and $(k,l)$, respectively. The value of these operators is equal to one if there is a dimer at the given bond and zero if there is no dimer. 
The reference dimer $(i,j)$ is shown by the thick black segments. The thickness of each segment scales with the magnitude of the correlation. Black (orange) segments denote positive (negative) correlation values. 
The fluid-like behavior of the $Z_2$ phase at $x\!=\!0$ and the characteristic pattern of the diamond VBC at $x\!=\!1$ show up clearly in the correlation patterns of Fig.~\ref{fig:SimplifiedModelResults}\,(a). The VBC pattern can be seen, in particular, by focusing on the pattern of positive correlations (black segments).

The transition between the $Z_2$ liquid and the VBC phases can be diagnosed in the ED spectra (Fig.~\ref{fig:SimplifiedModelResults}\,(b)) by the level crossing in the first excitation above the ground state. In the liquid region, the first excitation has momentum zero, but belongs to different topological sector from that of the ground state, while in the diamond VBC region, the first excitation has a finite momentum, consistent with the translational symmetry breaking of the crystal. 
The critical point can be seen by the opening of the extrapolated topological gap in the GFMC data of Fig.~\ref{fig:SimplifiedModelResults}\,(c), and is located at $x_c\!\simeq\!0.35$. 
Note that this value is far below the crossing between the two excitations of Fig.~\ref{fig:SimplifiedModelResults}\,(b), signifying a very large correlation length. We shall return to this important aspect below.

\vspace*{-0.3cm}
\subsection{Physical origin of the diamond VBC}\vspace*{-0.3cm} 
The cartoon picture of Fig.~\ref{fig:DomainWalls}\,(a) shows that the `8a-00' tunneling events play a central role in stabilizing the diamond VBC. 
Clearly, the `8c' processes are not important because their amplitudes are generally weaker compared to those of `8a' and `8b', see Table~\ref{tab:tR2+E}. 
The `8b' processes have a finite loop-density in the ground state but otherwise do not play a decisive role, even though their amplitude is slightly stronger than that of `8a-00'. The reason is that states involving `8b' loops necessarily involve a finite number of extra defect triangles, i.e. a finite density for `8b-01' and `8b-11', whose tunneling amplitude is weaker compared to that of `8a-00'. Consider, for example, the state shown in Fig.~\ref{fig:DomainWalls}\,(b), which contains both `8a' and `8b' loops, and where we have highlighted three nearby `8b' loops. If a tunneling event is taking place in the upper left or upper right loops, then the triangles connecting to the middle loop are empty half of the time. As a result, the middle `8b' loop enters the `8b-01' or `8b-11' configurations. 
So, although $|t_{8b\text{-}00}|$ is slightly stronger than $|t_{8a\text{-}00}|$, the maximum possible frequency of `8b-00' events is effectively smaller than that of `8a-00' events, due to phase space constraints.

While the ground state energy of $\mc{H}_{\text{m}}(x)$ is also affected by fluctuations, the above qualitative arguments establish that the origin of the diamond VBC is directly related to $x$, i.e. the indentations in the tunneling amplitudes across the nine (minimal) loop-8 events of Fig.~\ref{fig:L8}. 
This mechanism is qualitatively different from the ones reported previously, which involve either a strong negative $V_8$,~\cite{Hao14} or other potential terms (called $K$ in Refs.~[\onlinecite{Wan2013}] and [\onlinecite{Hwang2015}]). Such large potential terms cannot be justified from the microscopic side.~\cite{IoannisZ2}

\vspace*{-0.3cm}
\subsection{Domain walls and vortices in the minimal model}
\begin{figure}[!b]
\vspace{-0.2cm}
\includegraphics[width=0.49\textwidth,clip]{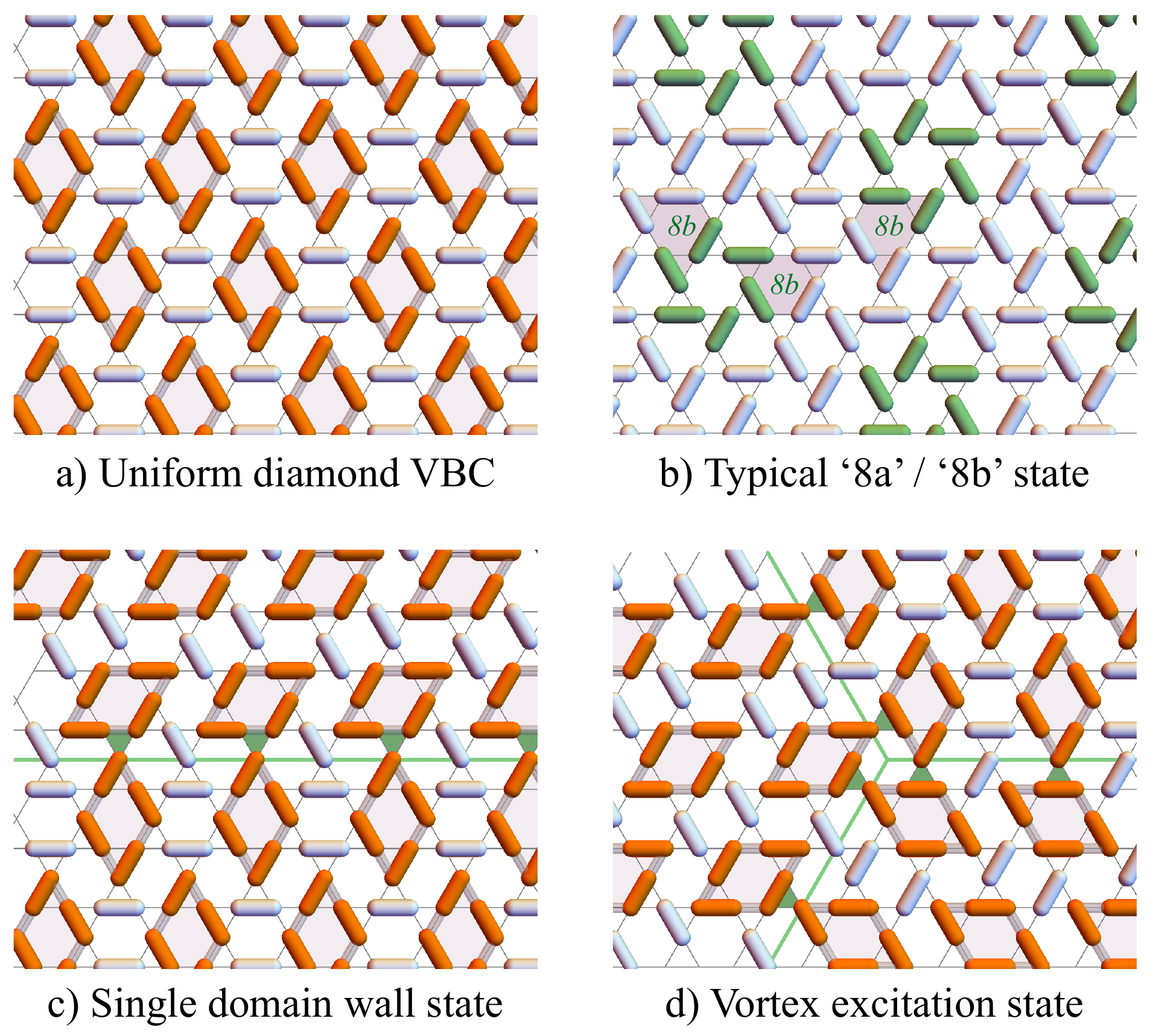}
\vspace{-0.45cm}
\caption{(a) Uniform diamond VBC state. (b) State with finite density of `8b' processes (2-colored loops). (c) A domain wall separates two different diamond VBC domains. (d) A vortex excitation corresponds to the intersection of three domain walls. The shaded triangles in (c-d) are the sources of the energy cost of the walls.}\label{fig:DomainWalls}
\end{figure}

\vspace*{-0.3cm}
Let us consider a domain wall excitation above the diamond VBC, see cartoon picture in Fig.~\ref{fig:DomainWalls}\,(c).  
In this state, all diamonds are again of the `8a' type, but unlike the uniform VBC, not all diamonds are of the `8a-00' type. Specifically, each of the vertical diamonds right below the domain wall are 50\% of the time in the configuration `8a-00' and 50\% in the configuration `8a-01'.  This is because one of the two NNVB states involved in the resonance of the horizontal diamonds right above the domain wall leaves a defect triangle below, see shaded blue triangles of Fig.~\ref{fig:DomainWalls}\,(b). These triangles are therefore the sources of the energy cost of the domain wall. This cost scales with the difference between the amplitudes of `8a-00' and `8a-01', i.e.\! it is proportional to $x$. We can also identify vortex excitations like the ones shown in Fig.~\ref{fig:DomainWalls}\,(d), which correspond to the intersection of three domain-walls. Clearly, the energy cost of these excitations scale also with $x$.

These considerations show that the melting of the VBC state at the critical point proceeds via the condensation of domain walls and vortices.
Moreover, if the system is inside the VBC state but close to the critical point, there is an extensive (sub-extensive) number of low-lying excitations associated with vortices (domain walls). This would be consistent with the large dimer correlation length and the high density of low-lying singlets found numerically.~\cite{Lecheminant1997,Waldtmann1998,Nakano2011,Laechli2011,Laeuchli2016}

\vspace*{-0.3cm}
\section{Effect of terms that are not included in the minimal model}\vspace*{-0.3cm}
\label{sec:NotIncludedTerms}
We now check how much of the above survives when we include the remaining terms of the `R2+E' model, starting from the loop-6 terms. 

\vspace*{-0.3cm}
\subsection{Loop-six processes}
\begin{figure}[!t] \vspace{-0.25cm}
\includegraphics[width=0.5\textwidth,clip]{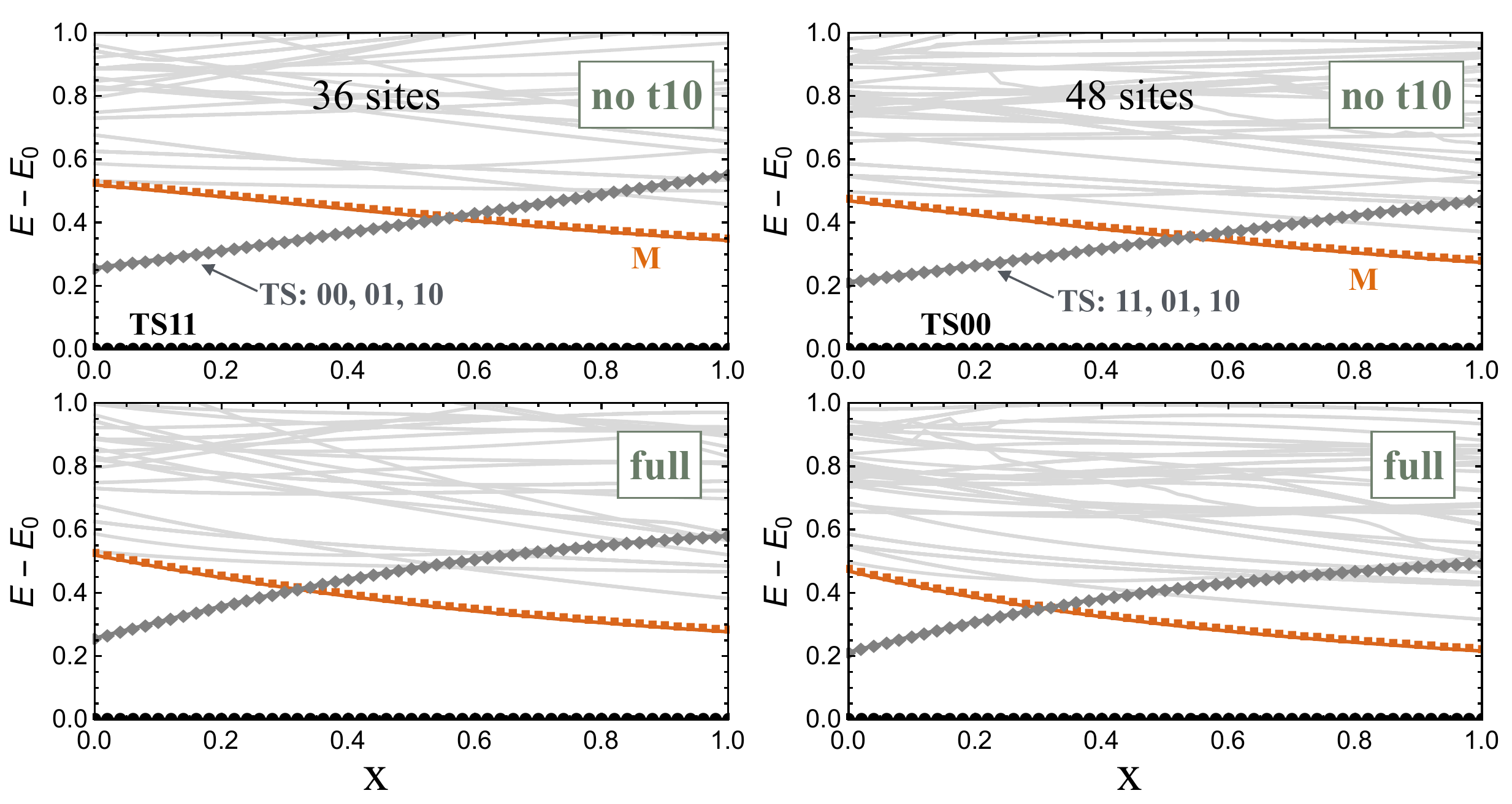}\vspace{-0.5cm}
\caption{Spectrum of $\mc{H}_{\text{m}}(x)$ with loop-6 (top), and loop-6 plus
loop-10 (bottom) processes added.}\label{fig:ExtraProcesses} 
\end{figure}

\vspace*{-0.3cm}
The loop-six terms are in fact larger than the loop-8 terms, which raises a legitimate concern.
Figure~\ref{fig:ExtraProcesses} (top) shows the low-energy spectra of $\mc{H}_{\text{m}}(x)$ with the addition of $t_6\!=\!0.127$ and $V_6\!=\!0.097$. The differences from Fig.~\ref{fig:SimplifiedModelResults}\,(b) are extremely small. So we can safely conclude that $t_6$ and $V_6$ do not play any appreciable role, despite being the largest in the `R2+E' model. 

This remarkable simplification is related to phase space constraints. There are $N_{\text{dc}}\!=\!2^{N/3+1}$ NNVB states for a torus geometry. Averaging over these states gives the following probabilities of a hexagon being in any given loop configuration: $p_6\!=\!p_{12}\!=\!\frac{1}{32}$, $p_{8a}\!=\!p_{10a}\!=\!\frac{3}{32}$, and $p_{8b}\!=\!p_{8c}\!=\!p_{10b}\!=\!p_{10c}\!=\!\frac{6}{32}$. Namely, loop-six (and loop-twelve) configurations are 15 times rarer than loop-eight or loop-ten. So, unless $t_6$ exceeds a high threshold (as e.g.\! at $R\!=\!0$~\cite{Schwandt2010,IoannisZ2}), loop-six processes are irrelevant, see also Ref.~[\onlinecite{AndreasQDM}].

\begin{figure*}[!t]
\includegraphics[width=0.33\textwidth,clip]{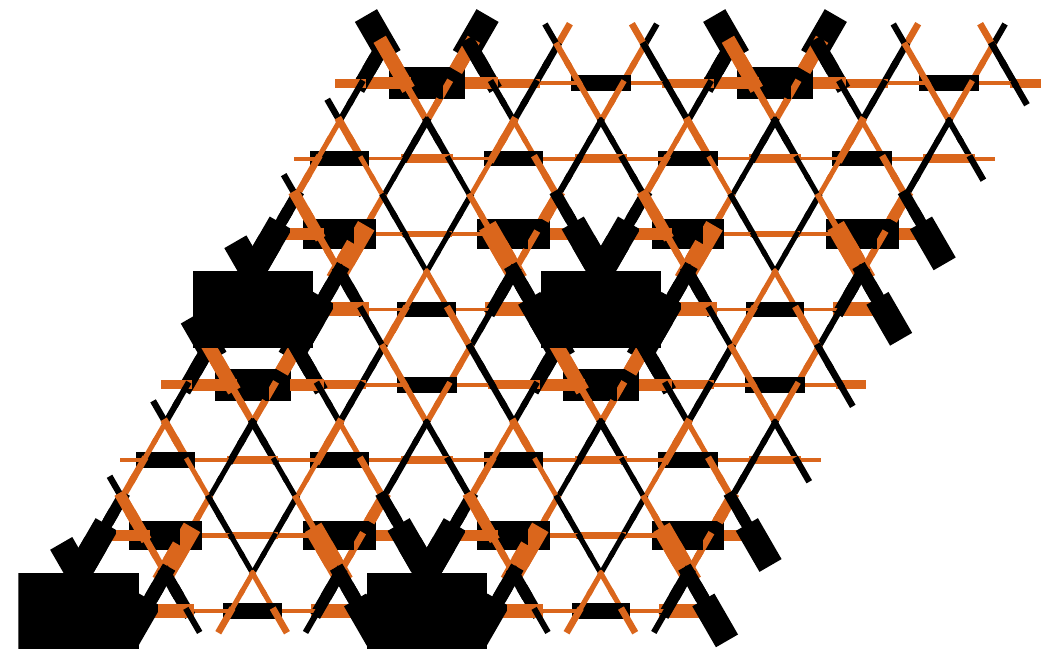}
\includegraphics[width=0.33\textwidth,clip]{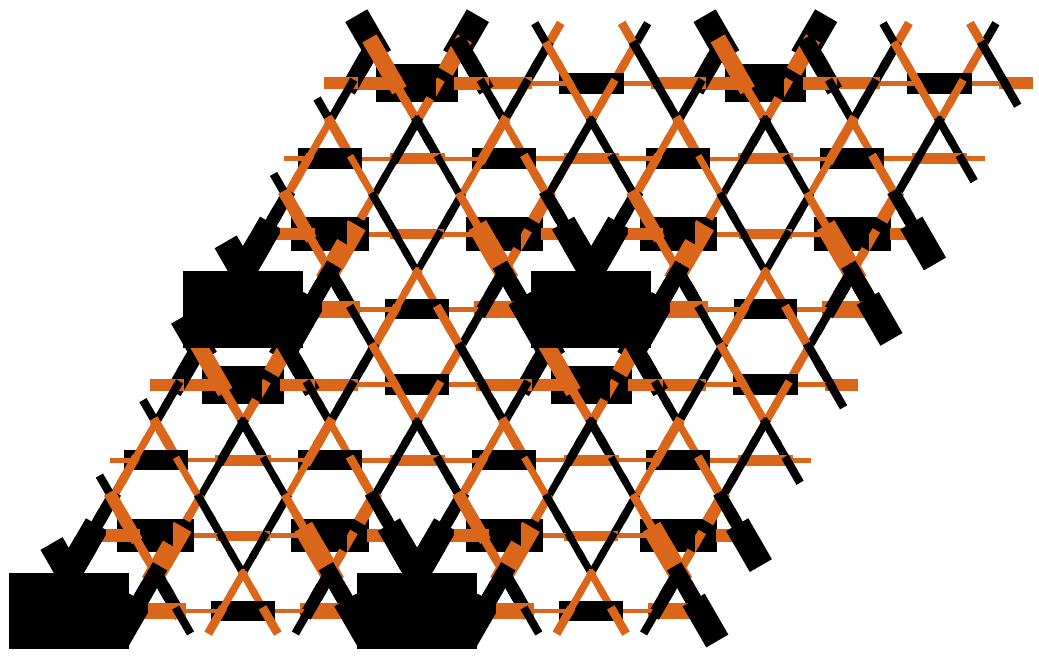}
\includegraphics[width=0.33\textwidth,clip]{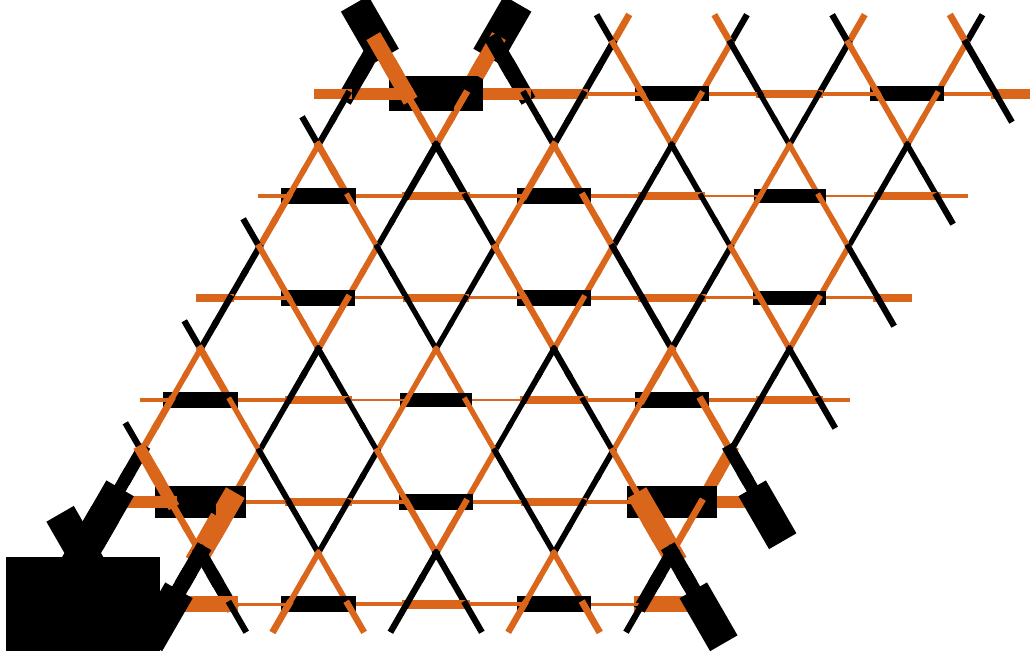}
\caption{Dimer-dimer correlations for the `R2+E' model with loop-ten processes excluded on the 48-site cluster (left), the full 'R2+E' model on the 48-site cluster (middle), and the full `R2+E' model on the 108-site cluster (right). The thick black segment at the bottom-left corner denotes the reference dimer. The 48-site cluster is periodically repeated four times for convenience. Black (orange) color indicates positive (negative) correlation values. The thickness of each dimer scales with the magnitude of the correlation. The diamond VBC order can be seen by the pattern of the positive correlations (black segments).}\label{fig:supp2}
\end{figure*}

\vspace*{-0.3cm}
\subsection{Loop-ten processes}\vspace*{-0.3cm}
Next come the loop-10 events. These are generally 1.5-2 times weaker than loop-8 but, unlike the loop-6, they appear equally often  with loop-8, as mentioned above. 
According to our numerics for the low-energy spectra (Fig.~\ref{fig:ExtraProcesses}, bottom panels) and the connected dimer-dimer correlations (Fig.~\ref{fig:supp2}), the ground state is the diamond VBC whether we include the loop-10 processes or not. In other words, these events do not give rise to another instability as long as their magnitude is in the range extracted by the cluster ED method. Nevertheless, the loop-10 processes take us slightly further away from the $Z_2$ spin liquid, because they have opposite signs from $t_8$ (compare e.g.\! with the integrable model of Misguich~\cite{Misguich02}).

\vspace*{-0.3cm}
\section{Ground state energy}
\label{sec:GSEnergy}
\vspace*{-0.3cm}
We will now show that the ground state energy of the `R2+E' model is fully consistent with published ED data for the original Heisenberg model, on 36-site~\cite{LeungElser1993,Waldtmann1998} and 48-site clusters.~\cite{Laeuchli2016} 
This analysis will further reveal that only about 5\% of the energy stems from tunneling. The remaining 95\% arises from potential energy contributions, that are practically the same for all NNVB states. This highlights why there are so many different orders that compete at tiny energy scales. 

The ground state energies of the Heisenberg model on the 36-site and 48-site clusters are $E^{(36)}_{\text{Heis}}/36\!=\!-0.438377J$~\cite{LeungElser1993,Waldtmann1998} and $E^{(48)}_{\text{Heis}}/48\!=\!-0.438703J$,~\cite{Laeuchli2016} respectively.
To extract the corresponding energies from the `R2+E' model, we must incorporate a global constant $c$ which is put aside when we go from the Heisenberg model to the effective QDM. 
The first main contribution to $c$ is the total energy of the singlets, $c_1\!=\!-\frac{3}{8}JN$. The second, $c_2$, comes from the potential energy of single defect triangles, which has been discussed in the Supplementing Material (Section D1) of Ref.~[\onlinecite{IoannisZ2}]. This contribution depends on the environment of the defect triangle and is in the range $-(0.291\!\pm\!0.025)JN$. There are $N/6$ defect triangles (the same for all NNVB states) which gives $c_2\!\simeq\!-(0.048\!\pm\!0.004)JN$. 
There are also contributions which vary from one NNVB state to the next, which are however much weaker and were not included in the `R2+E' model. These include the binding energy between two defect triangles, or the potential energies of the various processes other than $V_6$, which are both of the order of $0.01J$ or smaller.~\cite{IoannisZ2} 
There are also the corrections to the tunneling parameters from longer-range environments beyond $R\!=\!2$.~\cite{IoannisZ2} 
Furthermore, the Heisenberg energies contain contributions from tunneling loops that wind around the boundary, which are not included in the effective model explicitly.  
Altogether, the contributions beyond $c_1$ and $c_2$ that are not included in the `R2+E' model give an uncertainty of the order of $0.01JN$.

Now, the ground state energies of the `R2+E' model are
$E_{36}^{\text{(QDM)}}\!=\!-3.2939t_6$ and
$E^{(48)}_{\text{QDM}}\!=\!-4.2676t_6$, where $t_6\!=\!0.127J$. This gives
\be
E_{\text{QDM}}+c_1+c_2\!\simeq\!(-0.43\!\pm\!0.01) J N
\ee 
for both clusters. The Heisenberg energies are within these ranges, which is very satisfactory given the effective nature of the RVB description.
 
According to the above, the contributions $c_1\!+\!c_2$ correspond to about 95\% of the total ground state energy, and only the remaining 5\% comes from the tunneling physics. This remarkably small contribution explains why there are so many different orders that compete at tiny energy scales, as testified more explicitly by the excitation spectra of Fig.~\ref{fig:SimplifiedModelResults}\,(b) and \ref{fig:ExtraProcesses}.

\vspace*{-0.3cm}
\section{Comparison with DMRG}
\label{sec:DMRG}
\vspace*{-0.3cm}
The diamond VBC state is commensurate with several cylinder clusters used in the DMRG study of Ref.~[\onlinecite{YanHuseWhite2011}], yet the DMRG results deliver the $Z_2$ spin liquid and not the diamond VBC. 
To address this major concern we first return to an important observation made above. Namely, that the position of the level crossing between the two lowest excitations of Fig.~\ref{fig:SimplifiedModelResults}\,(b) is almost two times higher than the critical point $x_c\!\simeq\!0.35$, extracted from Fig.~\ref{fig:SimplifiedModelResults}\,(c). This large difference shows that the stabilization of the diamond VBC requires much larger system sizes than the ones of Fig.~\ref{fig:SimplifiedModelResults}\,(b). 

\begin{figure}[!b]
\vspace{-0.25cm}
\includegraphics[width=0.45\textwidth,clip]{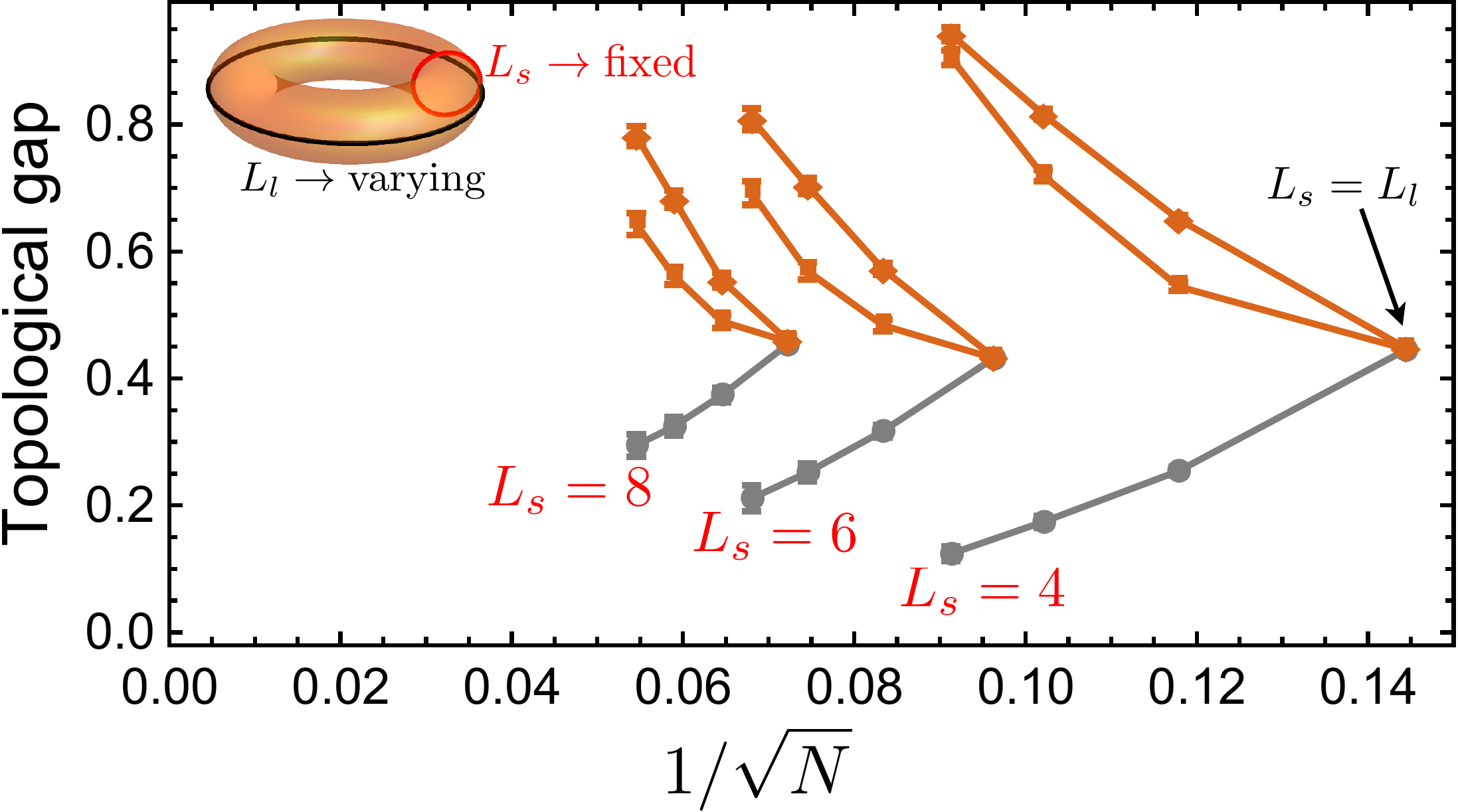}
\vspace{-0.25cm}
\caption{Topological gap (in units of $t_6\!=\!0.127J$) extracted from GFMC for the `T8+VS' model on tori with $L_s\!\times\!L_\ell$ unit cells. There are three sets of data with $L_s$ fixed and $L_\ell$ varying: 
$L_s\!=\!4$ (with $L_\ell\!=\!4$, $6$, $8$, $10$), 
$L_s\!=\!6$ (with $L_\ell\!=\!6$, $8$, $10$, $12$), 
and 
$L_s\!=\!8$ (with $L_\ell\!=\!8$, $10$, $12$, $14$), 
see App.~\ref{App:1} for the specific cluster geometries.}\label{fig:AssTorus}
\end{figure}

Importantly, the microscopic description allows one to check this crucial point explicitly at the level of $\mc{H}_{\text{m}}(x)$, which is free of the negative sign problem. To mimic the cylinder geometries of DMRG, we consider tori with $L_s\!\times\!L_\ell$ unit cells, and look at extrapolations with fixed $L_s$ and varying $L_\ell$. Figure~\ref{fig:AssTorus} shows three such extrapolations, with $L_s\!=$4, 6 and 8. 
The clusters with $L_s\!=\!6$ correspond to the largest circumference of 12 lattice spacings studied in Ref.~[\onlinecite{YanHuseWhite2011}].

Since we work on a torus we can still define four topological sectors, associated with the winding numbers $\{W_\ell,W_s\}$ along the directions of $L_\ell$ and $L_s$. For a symmetric torus with the three-fold rotation symmetry of the bulk, three out of the four topological sectors are degenerate. As soon as the torus becomes asymmetric, we lose the three-fold symmetry and each sector gives a different minimum energy. At the same time, we expect that only two levels (with the same $W_s$ but different $W_\ell$, see also App.~\ref{App:E}) approach each other with increasing $L_\ell$, if the $L_\ell\!\to\!\infty$ system is in the liquid phase. 
The GFMC results of Fig.~\ref{fig:AssTorus} confirm this general picture. 

More importantly, the extrapolation to the limit $L_\ell\!\to\!\infty$ gives the $Z_2$ liquid at $x\!=\!1$ for all three values of $L_s$ in Fig.~\ref{fig:AssTorus}. 
This is in stark contrast with extrapolations based on asymmetric tori with fixed aspect ratios (see App.~\ref{app:FAR}), which give results that are in perfect agreement with Fig.~\ref{fig:SimplifiedModelResults}\,(c).
So, for cylinder geometries the $Z_2$ spin liquid must eventually give way to the diamond  VBC for large enough $d$. The overall tendency in Fig.~\ref{fig:AssTorus}, with the extrapolated gap becoming less and less negative with increasing $L_s$ corroborates this picture. 
An upper boundary of the critical value is $L_s\!\sim\!10$ (i.e., 20 lattice spacings) because the largest cluster of Fig.~\ref{fig:SimplifiedModelResults}\,(c) has a linear size of 20.78 lattice spacings [incidentally, this is also above the 17 lattice spacings of Ref.~[\onlinecite{Shollwock2012}] (which are not commensurate with the diamond VBC)]. This is consistent with the fact that the clusters with $L_s\!=\!10$ are precisely the ones where GFMC fails to converge, because the liquid guiding wave-function~\cite{Binder2992,Sorella2000,Ralko2008} is not good any longer.

\vspace*{-0.3cm}
\section{Discussion}\label{sec:Disc}\vspace*{-0.3cm} 
To a large extent, the present study completes one of the oldest microscopic approaches to the spin-1/2 kagome problem.~\cite{ZengElser95,Mila1998,MambriniMila2000,Misguich02,Misguich03,Schwandt2010,Poilblanc2010,IoannisZ2,L4L8} The `R2+E' model incorporates the effects of virtual singlets up to the $R\!=\!2$ level (where all parameters except $t_6$ have essentially converged~\cite{IoannisZ2}), as well as the embedding dependence coming from the nearest triangles next to the loops. This model has 40 parameters that are all implemented in our numerics. In conjunction with the orthogonalization of the NNVB basis, this entails a microscopic description with unprecedented accuracy and sets the record in the program initiated many years ago by Rokhsar and Kivelson in their derivation of the effective RVB description of Heisenberg models.~\cite{RokhsarKivelson} 

Our numerical calculations of the microscopic QDM description place the system within the diamond VBC phase. The `distance' from the $Z_2$ spin liquid should be considered in relation to the fact that $x\!=\!1$ corresponds to about 30\% difference between $t_{8a\text{-}00}$ and $t_{8a\text{-}01}$, and the critical point $x_c\!\simeq\!0.35$ corresponds to about 10\%.  
Including fluctuations from the environment further away from the loop (which would increase the number of parameters significantly) will effectively give rise to a small reduction in the value of $x$ that quantifies the distance from the `T8' model. On the other hand, the weak, loop-10 processes act to effectively increase the distance from the liquid phase, as mentioned above. This entails a small uncertainty in the distance from the critical point. 

Despite this uncertainty, which reflects the non-local character of the virtual singlets, the microscopic tunneling description offers a simple and intuitive picture of several key aspects of the problem.
First, the 36-site VBC proposed in earlier works~\cite{MarstonZeng91,NikolicSenthil03,SinghHuse07,Schwandt2010,Poilblanc2010} is not one of the competing states~\cite{YanHuseWhite2011} because the shortest tunneling events that stabilize this state are irrelevant.  
Second, the diamond VBC state is one of the competing states~\cite{YanHuseWhite2011} because the second-shortest loops are the most relevant. In particular, if the ground state is described by short-range singlets then it must be the diamond VBC.  
Third, the $Z_2$ spin liquid becomes the ground state in cylinder geometries with small diameters $d$, consistent with DMRG.~\cite{YanHuseWhite2011,Shollwock2012,Jiang2012} 
This liquid is in fact adiabatically connected to  the integrable models of Misguich {\it et al}~\cite{Misguich02} and Hao {\it et al},~\cite{Hao14} despite the fact that these models appear very far in parameter space. 
Finally, the microscopic description also offers an interpretation for the high density of low-lying singlets found by ED,~\cite{Lecheminant1997,Waldtmann1998,Nakano2011,Laechli2011,Laeuchli2016} based on the overall tiny energy contribution from tunneling, and the presence of infinite domain wall and vortex states.

The dimer description leaves no room for a gapless U(1) liquid because every quantum dimer model on kagome can be mapped rigorously to a $Z_2$ gauge theory.~\cite{Wan2013,Hwang2015} 
So, in order to account for the evidence~\cite{Iqbal2013,Iqbal2014,He2016,Liao2017} that one of the competing phases is a U(1) spin liquid one must also examine the physics inside the orthogonal, longer-range singlet sector. 
Variationally, this sector onsets at an energy of $3J/2$ above the NNVB manifold, but longer-range singlets can lower their energy by the kinetic energy of the associated spinons.
Such a picture in terms of spinons interacting via an emergent U(1) gauge field has been proposed, although the strong attraction between spinons in the singlet channel seem to push the system into the $Z_2$ phase.~\cite{Hao2013} 
Extracting the microscopic parameters of this extended picture could give the right insights as to why the U(1) liquid is also one of the competing phases.

According to the above, if the ground state of the spin-1/2 kagome antiferromagnet is described by short-range singlets then it is a diamond valence bond crystal, but if the system is a quantum spin liquid, it has to involve long-range singlets, consistent with the algebraic liquid scenario.

\begin{figure}[!t]
\includegraphics[width=0.40\textwidth,clip]{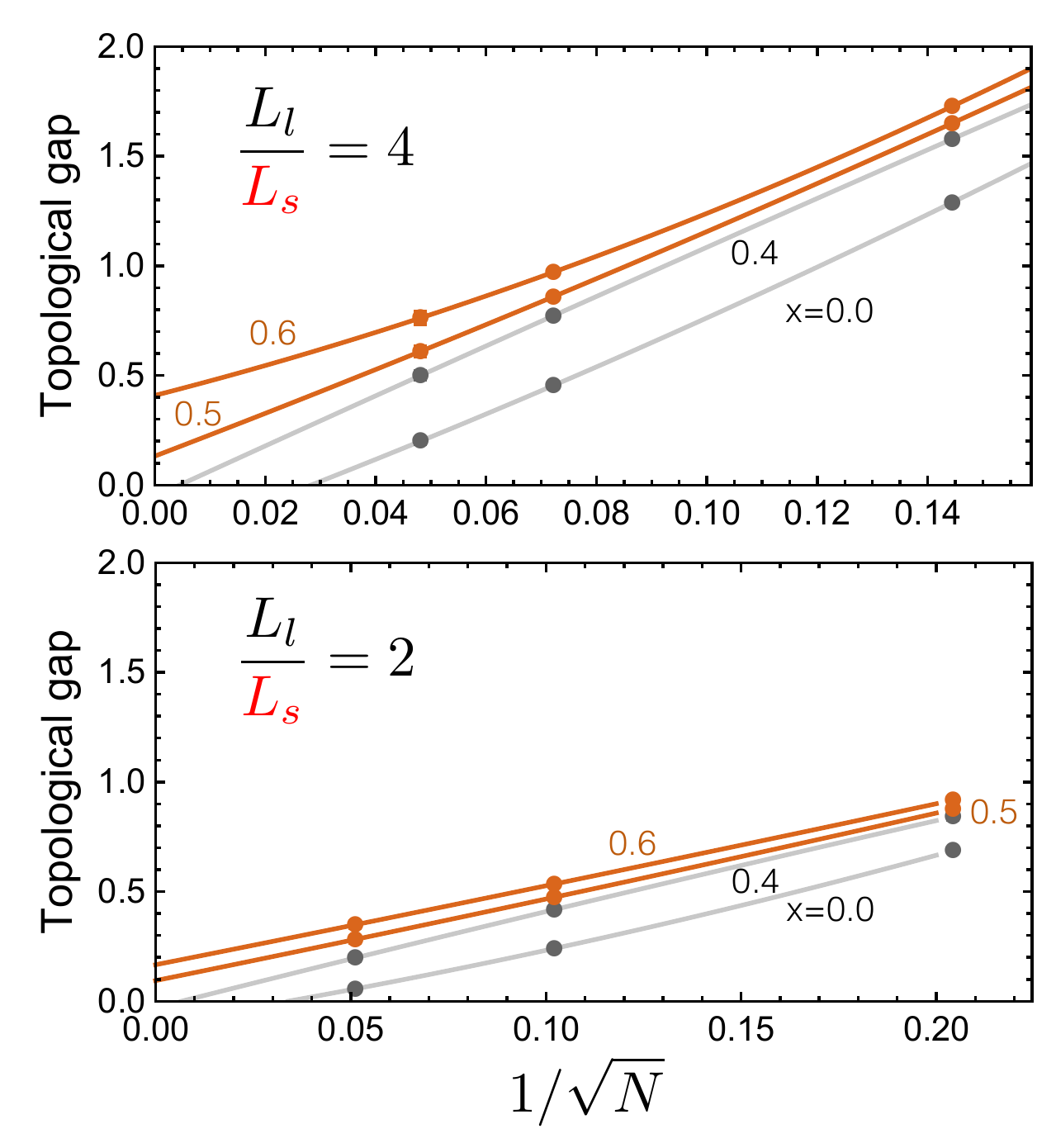}
\caption{Evolution of the topological gap as a function of $1/\sqrt{N}$, for different fixed ratios of $\frac{L_\ell}{L_s} = 4$ (top) and $2$ (bottom) and various $x$ in the QSL region (gray) and the diamond VBC region (orange).}\label{fig:TopGap2}
\end{figure}

\vspace*{0.2cm}
\noindent{\it Acknowledgments}. 
We acknowledge fruitful discussions with D.~Poilblanc, Y.~Iqbal, F.~Pollmann, Y.~Wan, N.~Perkins, M. D. Schulz and G.~Baskaran. We also acknowledge support from the Swiss National Science Foundation.
\vspace*{0.15cm}

\appendix

\vspace*{-0.5cm}
\section{Cluster geometries in our numerics of the effective QDM}\label{App:1}
\vspace*{-0.5cm}
The primitive vectors of the kagome lattice are ${\bf a}\!=\!{\bf x}$ and ${\bf b}\!=\!\frac{1}{2}{\bf x}\!+\!\frac{\sqrt{3}}{2}{\bf y}$. 
The spanning vectors of the symmetric torus clusters of e.g. Fig.~\ref{fig:SimplifiedModelResults} are ${\bf T}_1\!=\!-L{\bf a}\!+\!L{\bf b}$ and ${\bf T}_2\!=\!-L{\bf a}\!+\!2L{\bf b}$, which contain $3L^2$ unit cells and $N\!=\!9L^2$ sites. 
The spanning vectors of the asymmetric torus clusters of Fig.~\ref{fig:AssTorus} are ${\bf T}_\ell\!=\!L_{\ell}{\bf a}$ and ${\bf T}_{\text{s}}\!=\!L_{\text{s}}{\bf b}$. These clusters have $L_{\text{s}}L_\ell$ unit cells and $N\!=\!3L_{\text{s}} L_\ell$ sites. 
All clusters studied (symmetric and asymmetric) are commensurate with the diamond VBC state. 


\vspace*{-1cm}
\section{Finite size study of the topological gap for clusters with fixed aspect ratio $L_\ell/L_s$}\label{app:FAR}
\vspace*{-0.5cm}
Figure~\ref{fig:TopGap2} shows extrapolations of the topological gap based on asymmetric tori with fixed aspect ratios $L_\ell/L_s = 4$ (left) and $L_\ell/L_s=2$ (right). In both cases, a phase transition between the QSL (gray curves) and the diamond VBC (orange curves) occurs around $x\sim0.4$, consistent with the extrapolations based on regular clusters, in Fig.~\ref{fig:SimplifiedModelResults}. 

\begin{figure}[!b]
\includegraphics[width=0.3\textwidth,clip]{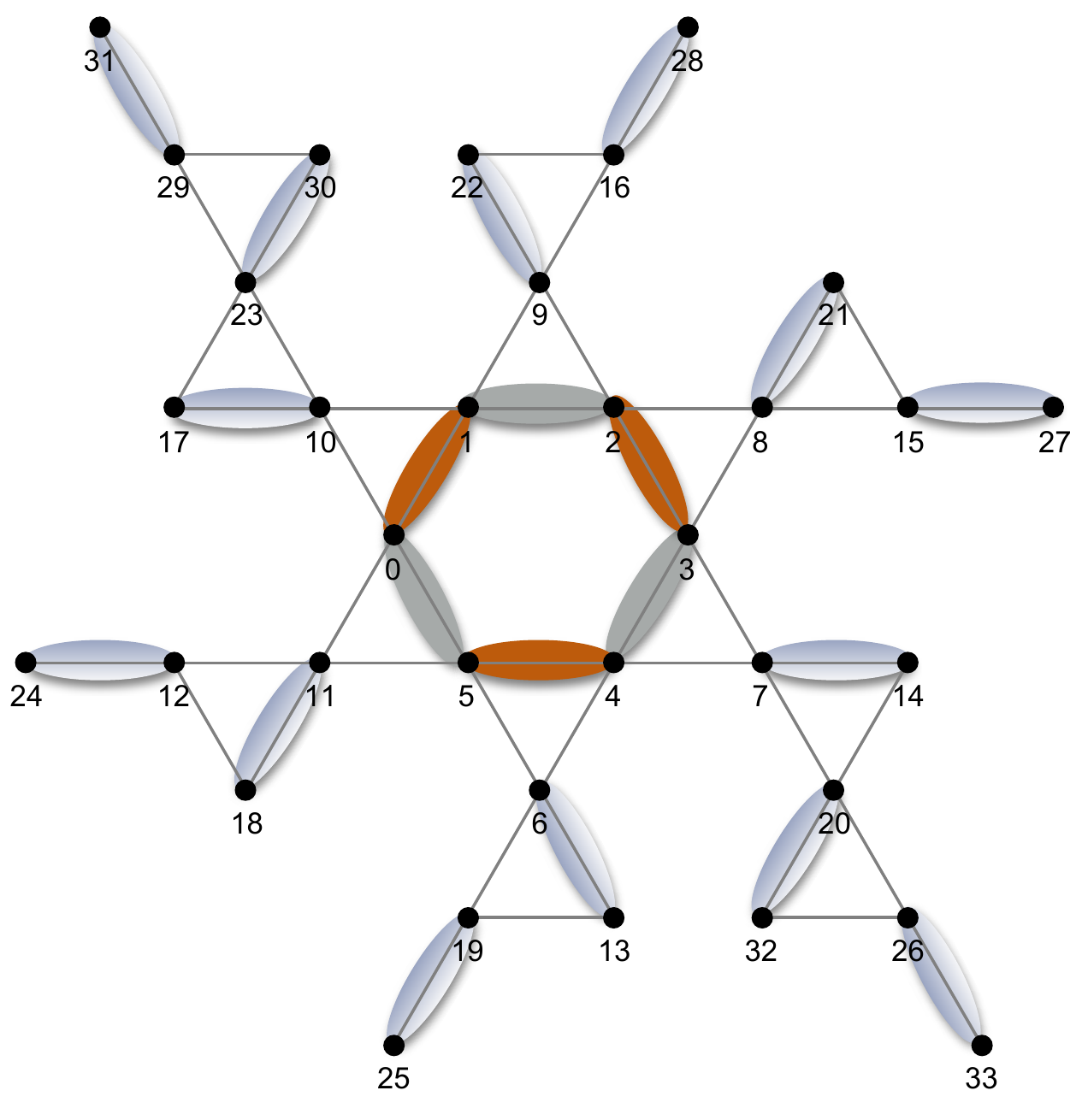}
\caption{The 34-site cluster used to extract the parameters $t_6\!=\!0.127J$ and $V_6\!=\!0.073J$ given in the main text.}\label{fig:HC34}
\end{figure}

\vspace*{-0.3cm}
\section{Heisenberg cluster used to extract the parameters $t_6$ and $V_6$}\label{app:HC6}
\vspace*{-0.5cm}
As shown in Fig.~4 of Ref.~[\onlinecite{IoannisZ2}], the tunneling parameter $t_6$ has not yet converged at the $R\!=\!2$ level. 
The cluster corresponding to $R\!=\!3$ has 42 sites and does not have enough symmetries to be treated by ED. 
The parameters $t_6\!=\!0.127J$ and $V_6\!=\!0.073J$ used here are extracted from the cluster shown in Fig.~\ref{fig:HC34}, which is intermediate between $R\!=\!2$ and $R\!=\!3$. This cluster contains 34 sites.
\vspace*{-0.5cm}
\begin{figure}[!t]
\vspace*{-0.5cm}
\includegraphics[width=0.45\textwidth,clip]{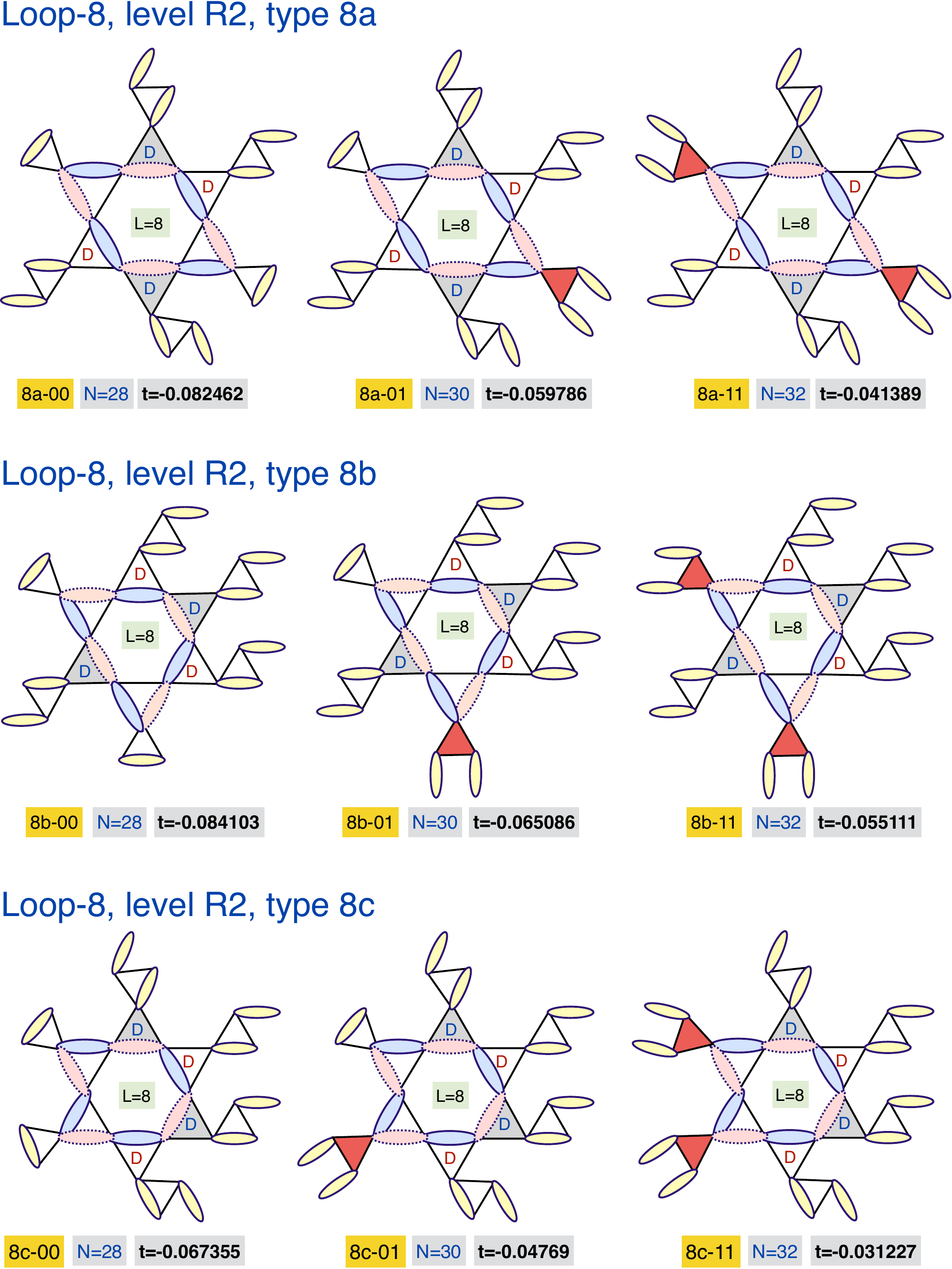}
\caption{Heisenberg clusters used to extract the tunneling amplitudes of the type `8a', `8b' and `8c' processes.}\label{fig:loop8}
\vspace*{-0.5cm}
\end{figure}
%
%
\begin{figure}[!t]
\vspace*{-0.5cm}
\includegraphics[width=0.45\textwidth,clip]{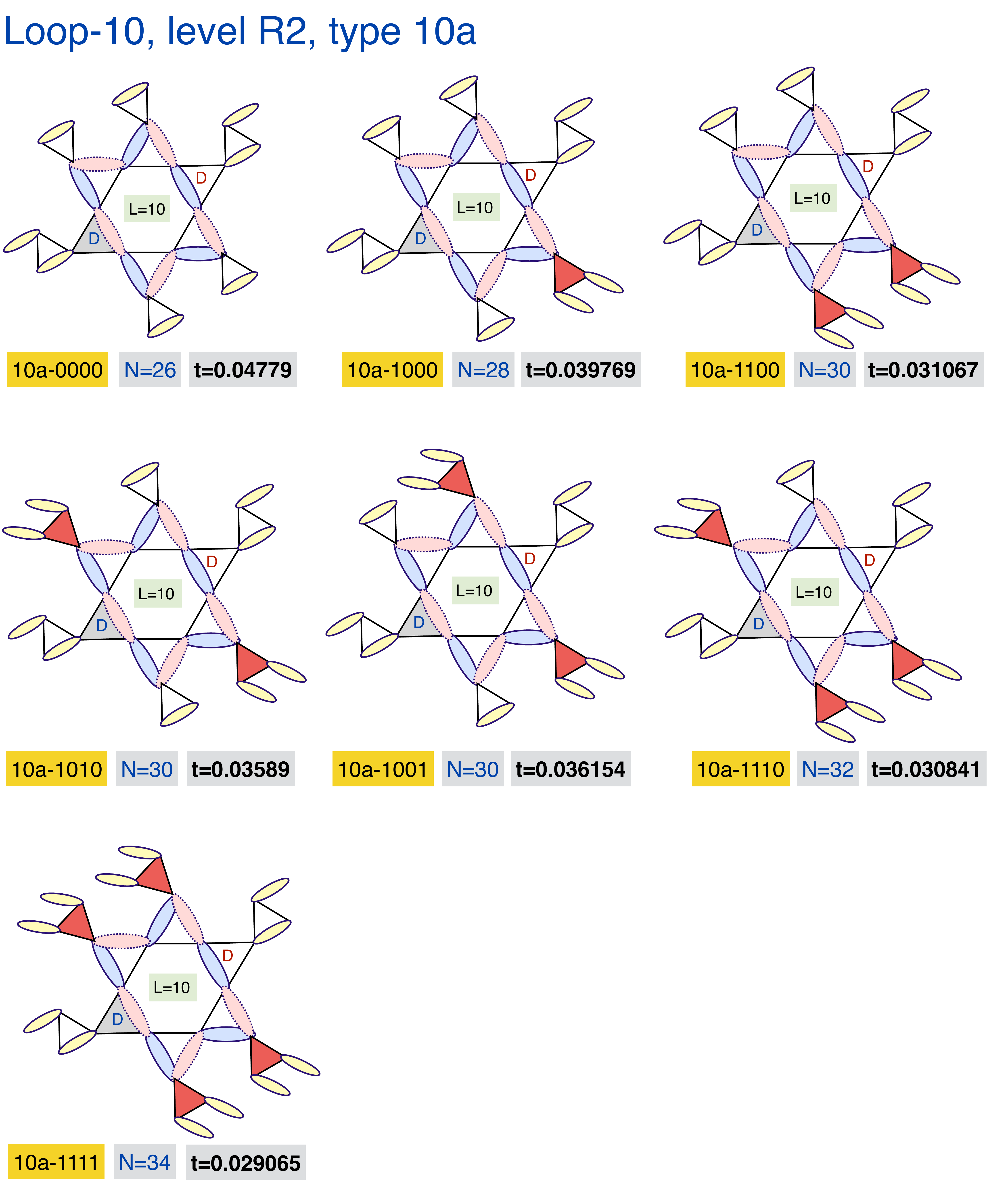}
\caption{Heisenberg clusters used to extract the tunneling amplitudes of the type `10a' processes.}\label{fig:loop10a}
\vspace*{-0.5cm}
\end{figure}
\vspace*{-0.5cm}
\begin{figure}[!t]
\vspace*{-0.5cm}
\includegraphics[width=0.45\textwidth,clip]{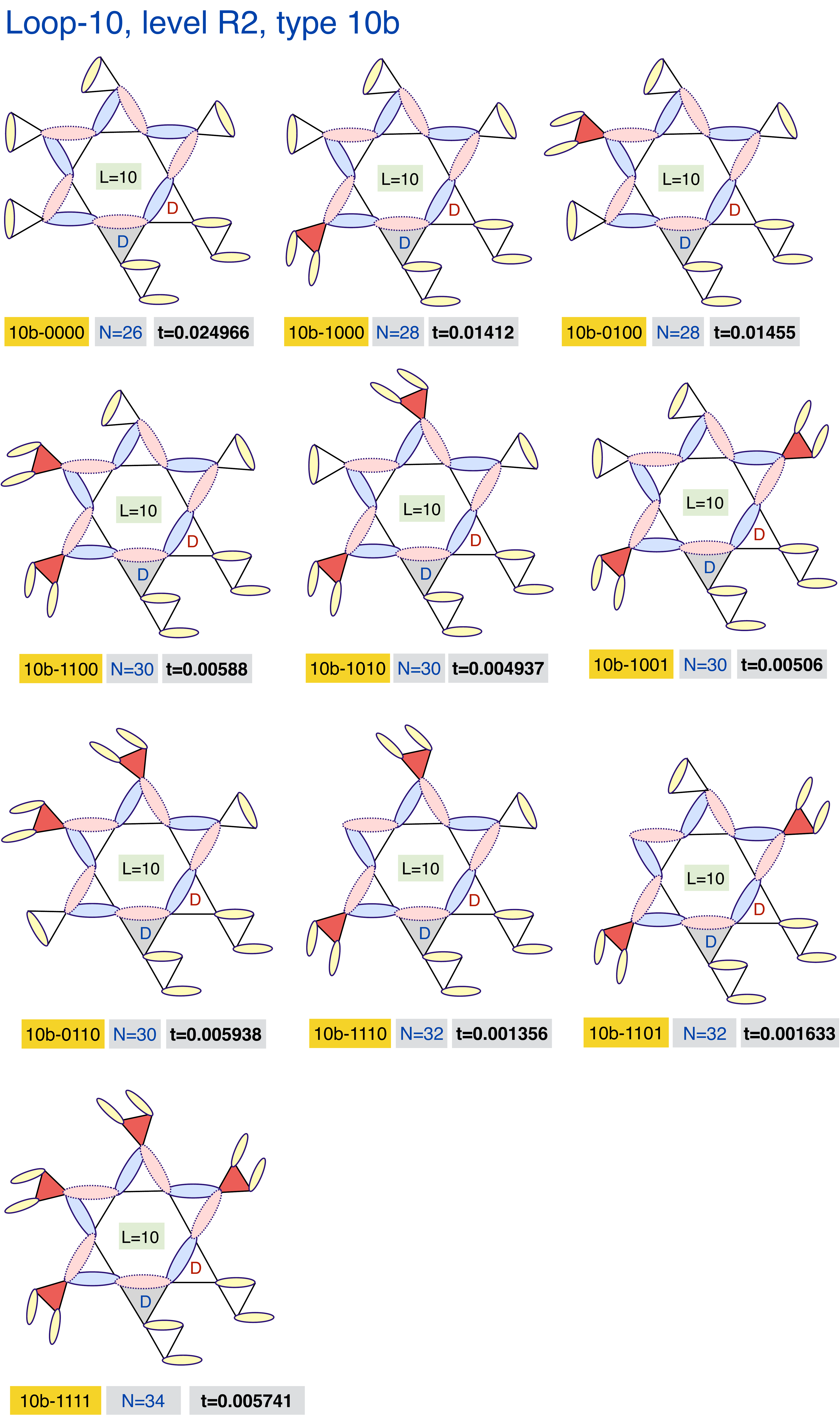}
\caption{Heisenberg clusters used to extract the tunneling amplitudes of the type `10b' processes.}\label{fig:loop10b}
\vspace*{-0.5cm}
\end{figure}

\section{Heisenberg clusters used to extract the\\
loop-8 and loop-10 amplitudes}\label{app:HC810}
\begin{figure}[!t]
\includegraphics[width=0.45\textwidth,clip]{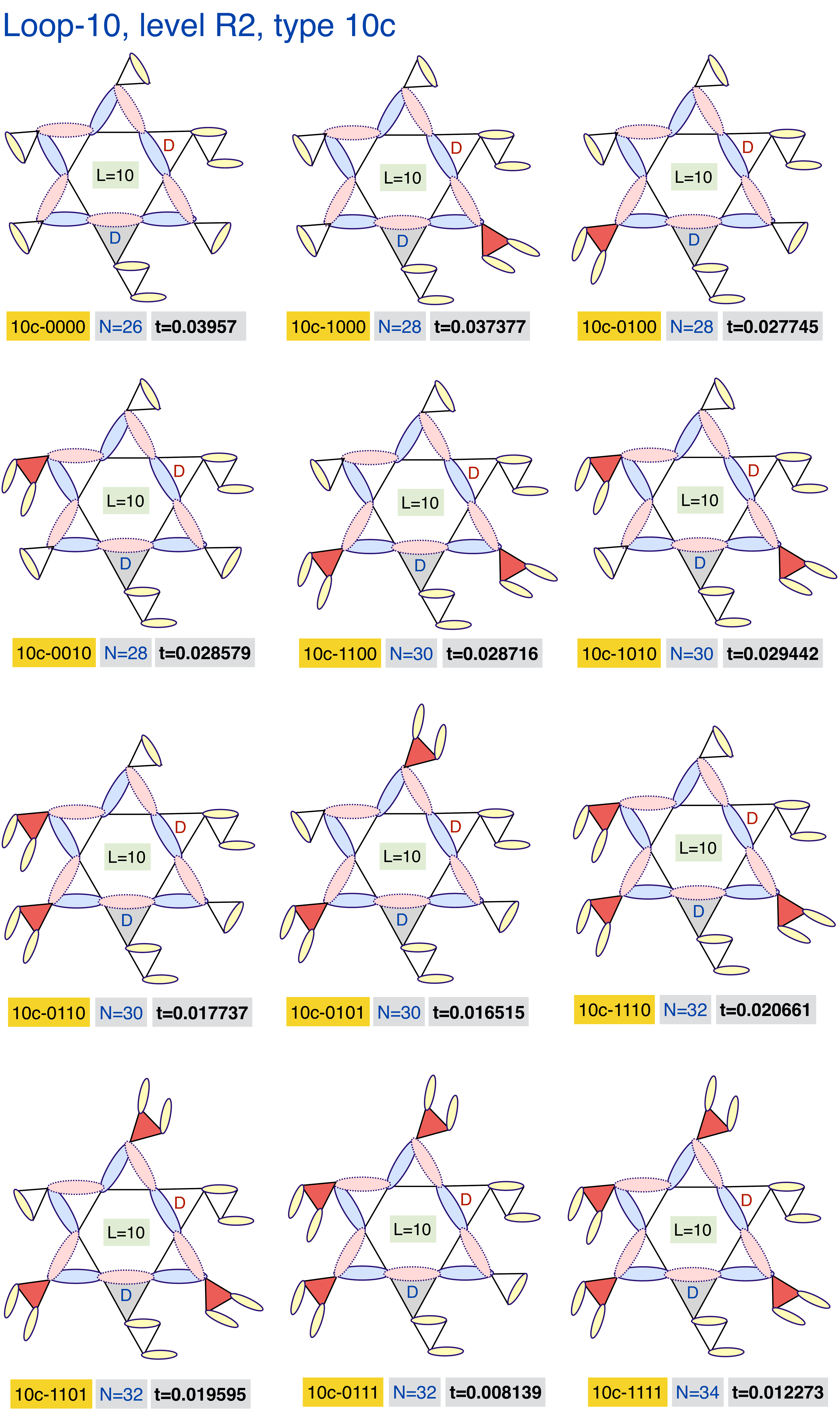}
\caption{Heisenberg clusters used to extract the tunneling amplitudes of the type `10c' processes.}\label{fig:loop10c}
\end{figure}

\vspace*{-0.5cm}
The Heisenberg clusters used to extract the tunneling amplitudes of the loop-8 and loop-10 processes are shown in Figs.~\ref{fig:loop8}, \ref{fig:loop10a}, \ref{fig:loop10b} and \ref{fig:loop10c}. The procedure to extract the parameters from the Heisenberg spectra of these clusters is described in detail in Ref.~[\onlinecite{IoannisZ2}].
The clusters are designed in such a way that they can accommodate only the two NNVB states that are involved in each given tunneling process that we are after.
These two NNVB states differ in the valence bond configuration along the central loop of length $L$, and are indicated by the blue (solid) and red (dashed) ovals. The yellow ovals denote the valence bonds away from the loop, which are common in the two NNVB states involved in the tunneling. 
The blue (red) letters `D' denote the positions of the defect triangles (triangles without dimers) when the dimers along the loops sit on the blue (red) ovals.
The shaded red triangles denote the `extra' defect triangles that appear in the nearby environment of the loop.
In all figures, for each cluster we provide the name of the process (as it appears in Table I of the main text), the number of sites of the cluster, and the tunneling amplitude $t$ extracted from the low-lying, tunnel split levels with the right symmetry, see details in Ref.~[\onlinecite{IoannisZ2}]. 
\vspace*{-0.5cm}
%
%
 

\begin{figure}[!b]
\subfigure[~Cylinder geometry.]{\includegraphics[width=0.4\textwidth,clip]{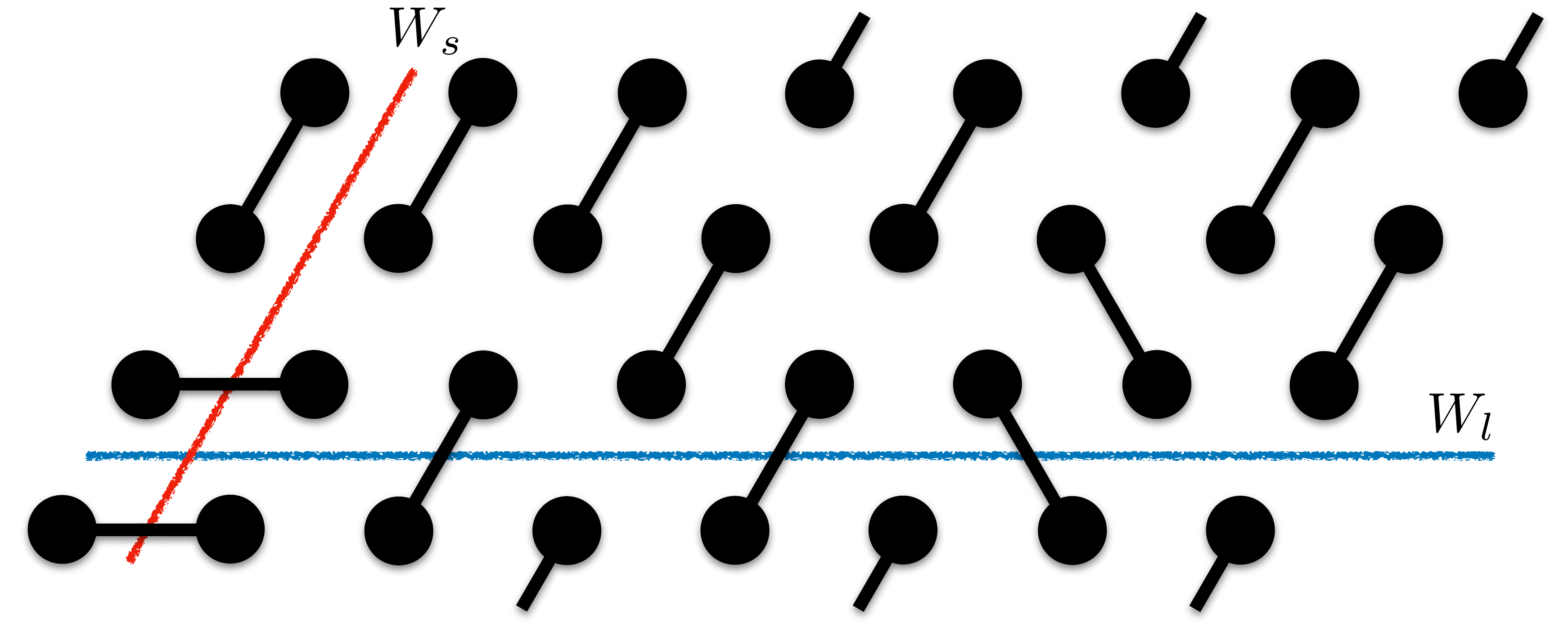}\label{fig:Cyl}}
\subfigure[~Topological gap.]{\includegraphics[width=0.4\textwidth,clip]{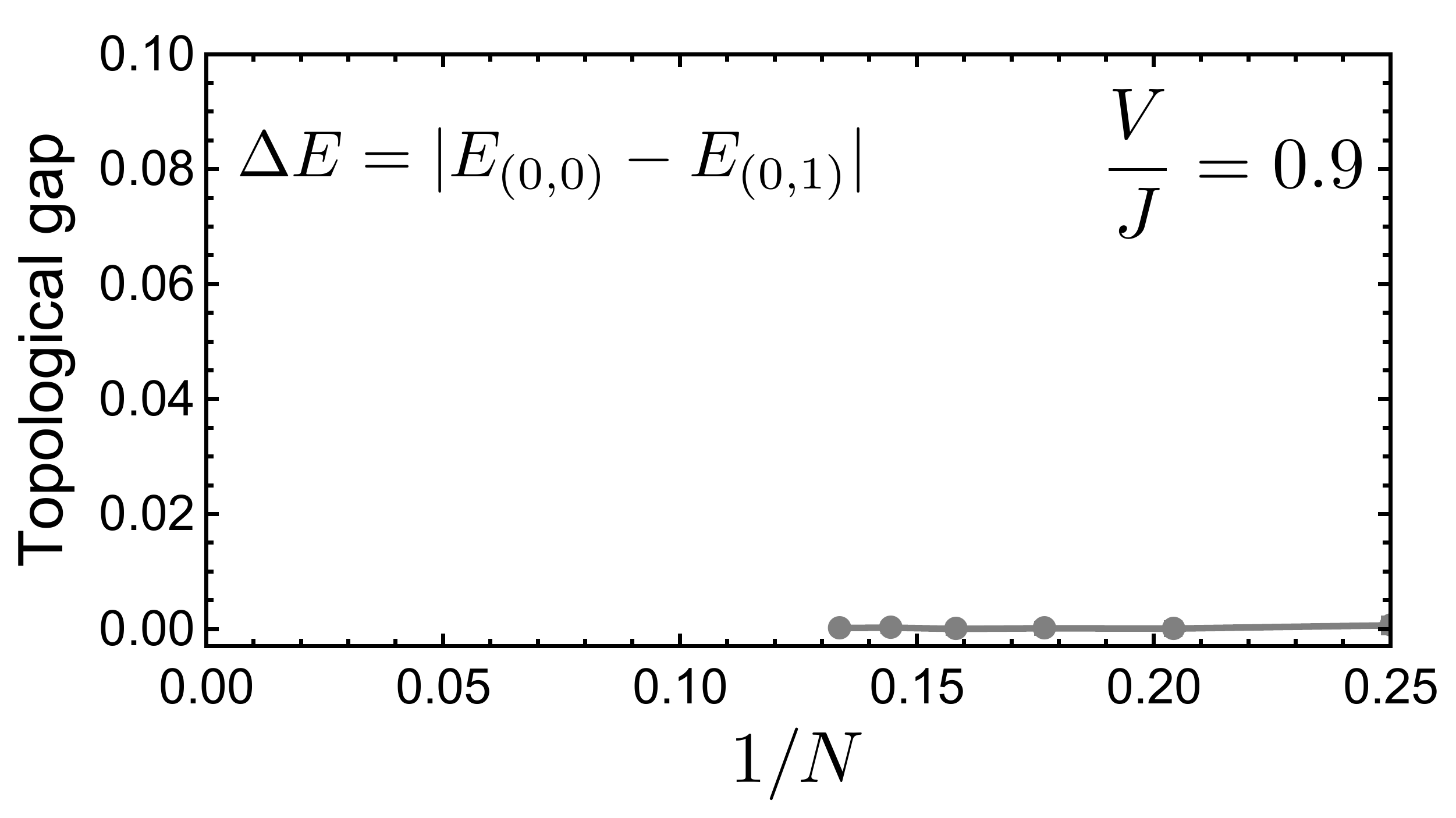}\label{fig:gapcyl}}
\caption{(a) Triangular cluster on a cylinder geometry with $(L_s,L_\ell)=(4,8)$ and open boundary conditions in the horizontal direction (blue) and periodic boundary conditions in the other direction (red).  The NNVB configuration shown lives in the topological sector $(W_s,W_\ell)=(1,0)$, where $W_s$ and $W_\ell$ are the parities of the numbers of dimers crossing the depicted cutlines (blue and red, respectively). (b) Topological gap between the sectors $(W_s,W_\ell)=(0,0)$ and $(0,1)$ on cylinders with $L_s=4$ and $L_\ell=4,6,8,10,12$ and $14$ for the triangular QDM with $V/J=0.9$.~\cite{Ralko2008}}
\end{figure}

\vspace*{-0.3cm}
\section{Cylinder geometry on the triangular lattice}\label{App:E} 
\vspace*{-0.3cm}
For a better understanding of which topological sectors are collapsing with the system size in the anisotropic torus geometry, we show numerical results on another well known model with a spin liquid ground state, the QDM model on the triangular lattice.~\cite{Ralko2008}
We consider cylinders with periodic boundary conditions (PBC) in one direction (red line), and open boundary conditions (OBC) in the other (blue line), as depicted in Fig.~\ref{fig:Cyl}.  
Since we have OBC along the horizontal direction, the topological winding number $W_s$ can only take a single value, which is equal to zero for the cylinders considered in Fig.~\ref{fig:Cyl}. On the contrary,  the winding number $W_\ell$ can take two possible values, 0 or 1. 

We have considered the vanilla quantum dimer model with only loop-4 processes containing a potential term (counting the number of flippable plaquettes) and a kinetic term (representing plaquette resonances) at ratio $V/J=0.9$, namely we are deep inside the RVB QSL phase.~\cite{Ralko2008} 
Our calculations for the topological gap are shown in Fig.~\ref{fig:gapcyl}. The energy difference between the two lowest energy levels is practically zero for fixed $L_s=4$ and $L_\ell=4,6,8,10, 12$ and $14$, while the two levels belong to the sectors $(W_s,W_\ell)=(0,0)$ and $(0,1)$. This shows that the two-fold topological degeneracy of the cylinder geometry involves states with different $W_\ell$ and not with the same $W_\ell$, consistent with the asymmetric kagome tori results of  Fig.~\ref{fig:AssTorus}.



%

\end{document}